\documentclass[twocolumn,tighten]{aastex62}
\pdfoutput=1 
\usepackage{amsmath,amstext}
\usepackage[T1]{fontenc}
\usepackage{ae,aecompl}
\usepackage[utf8]{inputenc}
\usepackage{apjfonts} 
\usepackage[figure,figure]{hypcap}
\usepackage{enumerate}
\usepackage{bm}
\DeclareGraphicsExtensions{}

\newcommand{\vpeak}{$V_{\rm peak}$}
\newcommand{\mpch}{Mpc $h^{-1}$}
\newcommand{\kms}{km s$^{-1}$}
\newcommand{\rsat}{$R_{\rm sub}$}
\newcommand{\esd}{$\Delta \Sigma$}

\shorttitle{SIDM in Cluster Substructure}
\shortauthors{Bhattacharyya et al.}

\begin{document}

\title{THE SIGNATURES OF SELF-INTERACTING DARK MATTER AND SUBHALO DISRUPTION ON CLUSTER SUBSTRUCTURE}

\email{bhattacharyya.37@buckeyemail.osu.edu}

\author[0000-0001-6442-5786]{Souradip~Bhattacharyya}
\affiliation{Presidency University, Department of Physics, 86/1 College Street, Kolkata 700073, India }
\affiliation{The Ohio State University, Department of Astronomy, Columbus, OH 43210, USA}
\author[0000-0002-0298-4432]{Susmita~Adhikari}

\affiliation{Department of Astronomy and Astrophysics, University of Chicago, Chicago, IL 60637, USA}
\affiliation{Kavli Institute for Cosmological Physics, University of Chicago, Chicago, IL 60637, USA}
\affiliation{Department of Physics, Stanford University, 382 Via Pueblo Mall, Stanford, CA 94305, USA}
\affiliation{Kavli Institute for Particle Astrophysics \& Cosmology, P. O. Box 2450, Stanford University, Stanford, CA 94305, USA}
\affiliation{SLAC National Accelerator Laboratory, Menlo Park, CA 94025, USA}

\author[0000-0002-5209-1173]{Arka~Banerjee}
\affiliation{Fermi National Accelerator Laboratory, Cosmic Physics Center, Batavia, IL 60510, USA}
\affiliation{Department of Physics, Stanford University, 382 Via Pueblo Mall, Stanford, CA 94305, USA}
\affiliation{Kavli Institute for Particle Astrophysics \& Cosmology, P. O. Box 2450, Stanford University, Stanford, CA 94305, USA}
\author[0000-0002-2986-2371]{Surhud More}
\affiliation{Inter-University Centre for Astronomy and Astrophysics, Post Bag 4, Ganeshkhind, Pune 411 007, India}
\affiliation{Kavli Institute for the Physics and Mathematics of the Universe (WPI), 5-1-5, Kashiwanoha, 277-8583, Japan}
\author{Amit~Kumar}
\affiliation{Inter-University Centre for Astronomy and Astrophysics, Post Bag 4, Ganeshkhind, Pune 411 007, India}
\author[0000-0002-1182-3825]{Ethan O.~Nadler}
\affiliation{Kavli Institute for Particle Astrophysics \& Cosmology, P. O. Box 2450, Stanford University, Stanford, CA 94305, USA}
\affiliation{Department of Physics, Stanford University, 382 Via Pueblo Mall, Stanford, CA 94305, USA}
\author{Suchetana~Chatterjee}
\affiliation{Presidency University, Department of Physics, 86/1 College Street, Kolkata 700073, India }

\begin{abstract}

The abundance, distribution and inner structure of satellites of galaxy clusters can be sensitive probes of the properties of dark matter. We run 30 cosmological zoom-in simulations with self-interacting dark matter (SIDM), with a velocity-dependent cross-section, to study the properties of subhalos within cluster-mass hosts. We find that the abundance of subhalos that survive in the SIDM simulations are suppressed relative to their cold dark matter (CDM) counterparts. Once the population of disrupted subhalos---which may host orphan galaxies---are taken into account,  satellite galaxy populations in CDM and SIDM models  can be reconciled. However, even in this case, the inner structure of subhalos are significantly different in the two dark matter models. We study the feasibility of using the weak lensing signal from the subhalo density profiles to distinguish between the cold and self-interacting dark matter while accounting for the potential contribution of orphan galaxies. We find that the  effects of self-interactions on the density profile of subhalos can appear degenerate with subhalo disruption in CDM, when orphans are accounted for. With current error bars from the Subaru Hyper Suprime-Cam Strategic Program, we find that subhalos in the outskirts of clusters (where disruption is less prevalent) can be used to constrain dark matter physics. In the future, the Vera C. Rubin Observatory Legacy Survey of Space and Time will give precise measurements of the weak lensing profile and can be used to constrain $\sigma_T/m$ at the $\sim 1$ cm$^2$ g$^{-1}$ level at $v\sim 2000$ km s$^{-1}$.


\end{abstract}

\keywords{dark matter}

\section{Introduction} \label{sec:intro}

The cold dark matter (CDM) paradigm, which treats the microphysical constituents of dark matter as collisionless, has been very successful in explaining the large scale structure of the Universe. Within this paradigm, N-body simulations have been used extensively to model the formation of nonlinear structure in the Universe \citep[e.g.,][]{2012PDU.....1...50K}. However, certain predictions derived from N-body simulations of CDM were thought to be in tension with observations of small-scale structure in the Universe, as inferred from dwarf galaxies \citep{2017ARA&A..55..343B}, particularly the core-cusp issue \citep[e.g.,][]{2004MNRAS.351..903G,2010AdAst2010E...5D} or the missing satellite problem \citep[e.g.,][]{Klypin9901240,Moore9907411}. Self-interacting dark matter (SIDM) was originally proposed as a viable candidate to mitigate some of these small-scale problems. In its most basic form this model allows dark matter particles to undergo elastic scattering, allowing for the exchange of momentum and energy via non-gravitational processes \citep[e.g.,][]{Spergel9909386,2000ApJ...534L.143B,2000ApJ...535L..21M}. SIDM models have a range of phenomenological signatures; namely allowing both for the formation of cores instead of cusps at the center of collapsed halos, and also a possible suppression of substructure, within massive hosts \citep{Tulin170502358}. 

With recent advances in observational precision, combined with more sophisticated numerical techniques, especially the ability to realistically model the effects of baryonic physics on these scales, some of these problems like the missing satellite problem has mainly been attributed to observational incompleteness \citep{Kim1812121,2020ApJ...893...48N} and the effects of reionization \citep{2000ApJ...542..535G,2002ApJ...572L..23S,2002MNRAS.333..177B}. However, to constrain the microphysics of dark matter, it remains important to study the effects of these properties, like self-interactions, on different observables from cosmological surveys. Depending on the particulars of the underlying model, these self-interactions can have a velocity-dependent differential cross section. This implies that such models are best constrained by combining information from systems which have different natural velocity scales. On one end of this spectrum are dwarf galaxies, which can probe interactions at very low relative velocities. At the other end of the spectrum are galaxy clusters inhabiting massive dark matter halos, which serve as laboratories to study some of the most energetic processes in the universe \citep{Markevitch:2003at, 2012ARA&A..50..353K}. These clusters are naturally the systems best suited to study effects of self-interactions at very high velocities. 

The gravitational potential of the dark matter halos around galaxy clusters can be studied in detail, both, through gravitational lensing observations and the spatial and velocity distribution of the population of satellite galaxies within them \citep[e.g.,][]{2004ApJ...617L..13N,2004ApJ...604...88S,2007MNRAS.376..180N,2013ApJ...765...24N}. Historically, galaxy clusters have been used to place constraints on dark matter cross-sections at the velocity dispersion scales set by the clusters' gravitational potential, i.e., $\sim 1000$ \kms{} \citep[e.g.,][]{2001ApJ...561...61G,2002ApJ...564...60M, 2004ApJ...606..819M, 2013MNRAS.430..105P, 2018MNRAS.474..746B,2019MNRAS.488.1572H}. Some of the recent constraints on $\sigma /m$ are $1$ cm$^2$ g$^{-1}$ at $\sim 3200$ \kms{}, as derived from merging clusters \citep{2017MNRAS.469.1414K} or $<0.5$ cm$^2$ g$^{-1}$ at $\sim 1150$ \kms{} estimated from lensing and stellar kinematics \citep{2020arXiv200612515S}. While most early studies of SIDM focused on velocity-independent differential cross-sections, various models with explicit velocity-dependence in the differential cross-section have been proposed in order to reconcile cluster constraints with lower bounds on the cross-section from smaller galaxies. Such velocity-dependent cross sections naturally arise in models where self-interactions are mediated by light particles \citep[e.g.,][]{Kaplinghat158003339}. 

In a universe where dark matter has a velocity-dependent interaction cross-section, the evolution of satellite galaxies, that live in their own dark matter halos, depend both on the interaction cross-section at the high velocity scale of the cluster's velocity dispersion and also the lower velocity scale of the subhalos' own internal velocity dispersion. Young objects like galaxy clusters, where a large fraction of their most massive subhalos and their satellite galaxies have not had enough time to get tidally disrupted, provide a unique opportunity to constrain the shape of the velocity dependence of the cross-section using a single system.
  
 In this paper, we study the population of subhalos in zoom-in simulations of 30 cluster-mass objects for a velocity dependent SIDM model. We use a relatively high normalization for the cross-section so that the effects of the self-interactions are prominent compared to the noise due to scatter in the masses and other properties of the zoom-in systems. The zoom-in method in particular, allows us to robustly simulate a wide range of scales that simultaneously encompass the massive host halo and its lower mass substructures.
 
 A host of ongoing and future surveys will provide us with large samples of galaxy clusters, allowing us to carry out statistical studies of the population of satellite galaxies in these systems. In particular, surveys like the Vera C. Rubin Observatory Legacy Survey of Space and Time (LSST) \citep{Abell:2009aa}, Dark Energy Survey (DES) \citep{Abbott:2005bi} and the Hyper-Suprime Cam Subaru Strategic Program (HSC-SSP) \citep{2018PASJ...70S...4A} will allow us to observe satellite galaxies that are, at least, a hundred times fainter than the Milky-Way, if not more. Meanwhile, the simulations allow us to make reliable predictions for the spatial and velocity distribution of the subhalos and their detailed internal structures. Understanding the evolution of these systems can therefore significantly boost our knowledge of dark matter microphysics in the near future \citep{2019BAAS...51c.207B,Drlica-Wagner190201055}.  
 
 The internal structures of satellites, in particular, can also help disentangle degeneracies between dark matter microphysics and baryonic effects. Comparing the matter distribution around galaxies with similar optical properties in clusters with those in fields can help factor out the baryonic effects on the galaxy's dark matter halo to a large extent. A widely used method to probe the structure of a galaxy's dark matter is through weak lensing, the ongoing HSC-SSP being a deep survey, is ideally suited for weak lensing studies of satellite galaxies. Using mock satellite distributions from our simulations and the covariance estimates of weak lensing measurements from the HSC survey \citep{KumarMoreinprep}, we explore the limits that a HSC-like survey can place on SIDM cross-sections using these observables. By rescaling the covariance matrix to acccount for various improvements, we also make predictions for a LSST-like survey.
 
 Importantly, this is the first study that accounts for the ``orphan'' galaxy population when making predictions for observations from simulations, both in CDM and SIDM. While comparing observed galaxy distributions to subhalo distributions from dark matter only N-body simulations, one must account for subhalos that have been artificially disrupted in the simulations or failed to be tracked due to numerical effects \citep{2001MNRAS.328..726S,2006MNRAS.371..537W,2018MNRAS.477..359C,VandenBosch171105276,VandenBosch180105427,2021arXiv210201837D}. The galaxies harbored by these subhalos, being more compact, may survive even after their corresponding subhalos disrupt. This effect has been known to bias subhalo and satellite population predictions if not properly accounted for in CDM simulations \citep{2004MNRAS.352L...1G}. In principle, these effects can be even more severe in SIDM, due to the additional evaporation of particles from subhalos due to self-interactions. Orphan modeling is therefore an important systematic to consider while forward modelling observations of satellite populations.

This paper is organized as follows. In Section \ref{sec:sims} we discuss the details of the simulations we use. In Section \ref{sec:subdistrib} we describe the subhalo distributions in clusters and in Section \ref{sec:sublensing} we describe the weak lensing analysis. The usage of $r$ will imply 3D distance compared to $R$ which will stand for the 2D projection of $r$.

\section{Simulations} \label{sec:sims}

To simulate self-interacting dark matter (SIDM) we use the method adopted in \cite{2013MNRAS.430...81R}, which modifies the evolution of dark matter particles in the GADGET-2  \citep{2005MNRAS.364.1105S} simulations, by introducing a Monte-Carlo particle scattering scheme. \cite{Banerjee190612026} extrapolates this scheme to include a velocity dependent scattering cross-section. In this paper, we extend this method to run zoom-in simulations of massive galaxy clusters that have been selected from a parent box of 1 Gpc $h^{-1}$ and $1024^3$ particles. We resimulate 30 clusters with virial mass between $(2.0-4.5) \times 10^{14}~ M_{\odot}~h^{-1}$ at redshift, $z=0$. The particle mass resolution in the zoom-in region is $m_p = 1.5 \times 10^8 ~M_{\odot}~h^{-1}$. We set the search radius for self-interactions equal to the gravitational softening scale, $h_{SI}=1$ kpc $h^{-1}$. These simulations were run with the cosmological parameters set to $\Omega_m$ = 0.3, $\Omega_{\Lambda}$= 0.7, $A_s$= 2.2$\times$10$^{-9}$, $n_s$= 0.96, $H_{0}$= 70 \kms{} Mpc$^{-1}$ and $\sigma_8$= 0.85 \citep{Banerjee190612026}. 

\begin{figure} \label{Figure:cross_sec}
\centering
\includegraphics[scale=0.35,trim={0 0.1cm 0 0},clip]{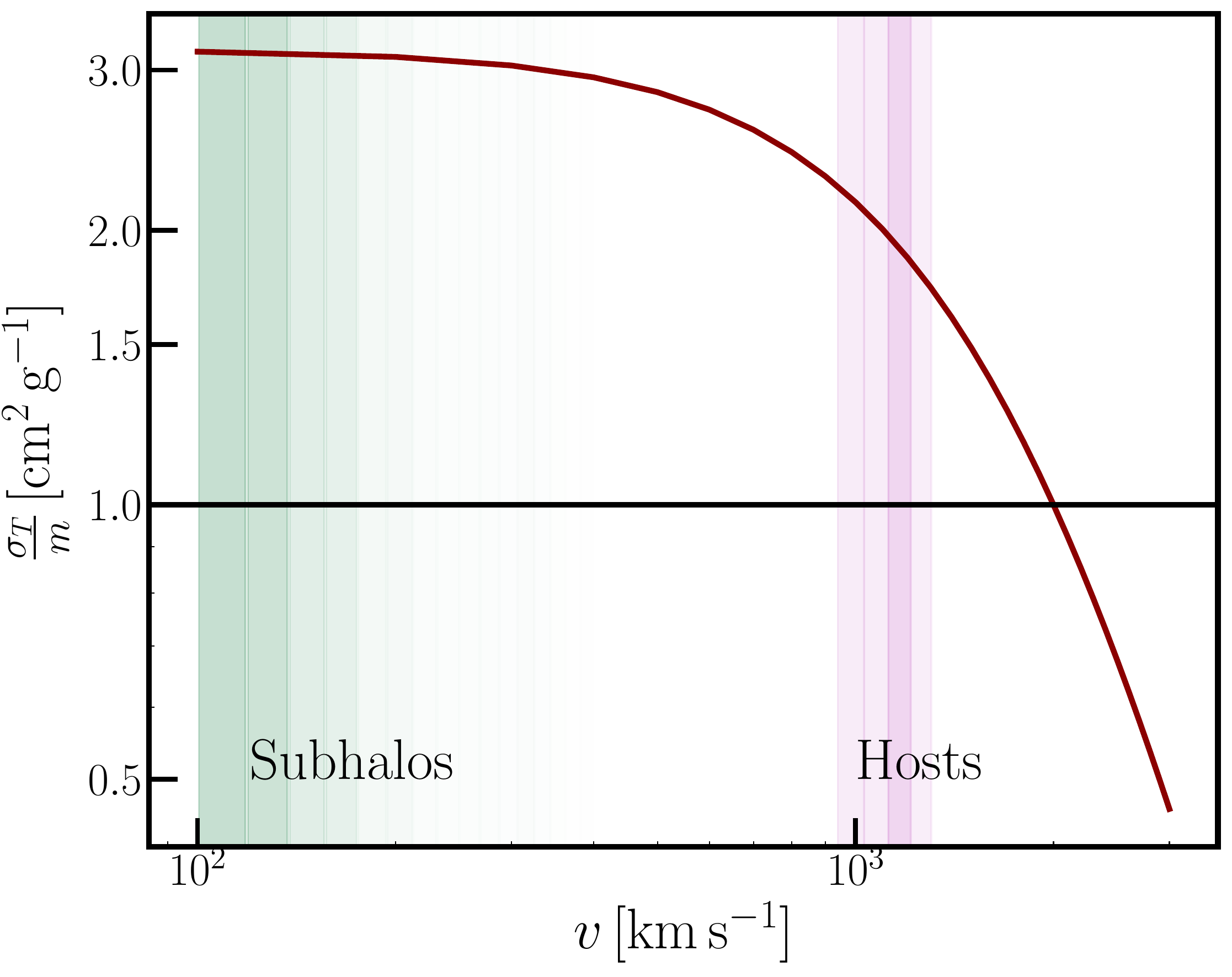}
\caption{The velocity dependence of the momentum transfer cross-section $\sigma_T /m$ for the SIDM model used in this paper. The green and pink vertical bars represent the distribution of velocity dispersion of the subhalos and their hosts respectively. Darker bands have higher numbers of objects. The relevant interaction cross-section between host particles and subhalo particles, which have relative velocities of the order of the host velocity dispersion is $\lesssim 2$ cm$^2$ g$^{-1}$ and decreases  in the Rutherford scattering limit. The relevant cross-section for the interaction between subhalo particles is at the subhalo velocity dispersion scale.}
\end{figure}

In this work we focus on a velocity dependent simulation cross-section where the differential cross-section also has an angular dependence \citep{Ibe09125425,2017MNRAS.467.4719R}. This is natural in models where dark matter interacts via \textit{dark photon} mediators, or in other words a Yukawa type of potential \citep{Kaplinghat158003339},
\begin{equation}
\frac{d\sigma}{d\Omega} = \frac{\sigma_0}{2\Bigg(1+\frac{v^2}{w^2}\sin^2\Big(\frac{\theta}{2}\Big)\Bigg)^2}.
\end{equation}
This model is parameterized by $w$, a characteristic velocity, below which the cross-section is isotropic as $\sigma \sim \sigma_0$ but above which the cross-section not only decreases with increasing velocity ($\propto v^{-4}$) but also becomes anisotropic, favouring scatterings by small angles \citep{Kahlhoefer:2013dca}. We use a value of $w = 1500$ \kms{} in this work as this corresponds to the upper limit of velocity scale of particles in a typical cluster sized halo. The normalization $\sigma_0$ is chosen such that the momentum transfer cross-section $\sigma_T /m$ at $v=2000$ \kms{} (the typical relative velocity between particles in the cluster) is $1$ cm$^2$~g$^{-1}$ \citep{Robertson:2016xjh, Markevitch:2003at}. This particular scale is chosen to approximately correspond to the constraints from the Bullet cluster which is $\sigma/m= 2$ cm$^2$/g (note $\sigma/m\sim 2\sigma_T$). The momentum transfer cross-section $\sigma_T/m$ shown in Fig.\ \ref{Figure:cross_sec} , is defined as \cite{Kahlhoefer:2013dca},

\begin{equation}
\sigma_T = \int \frac{d\sigma}{d\Omega} (1-\lvert\cos\theta\rvert) d\Omega.
\end{equation}

The velocity dispersion scales of the subhalos and their hosts as encountered in the simulations are shown as the green and pink bands respectively. We refer the reader to \cite{Banerjee190612026} for a detailed description of the simulation method.

\textit{Terminology:} The dark matter halo corresponding to the largest structure of the cluster will be referred to as the "cluster host'' or simply as the ``host.'' Following the terminology adopted in \citet{2020ApJ...896..112N}, smaller halos that have been accreted on to such a host will be divided into two categories:
\begin{itemize}
    \item \textit{Surviving subhalos} represent halos identified by {\sc Rockstar} in the zoom-in simulations at $z = 0$. We refer to galaxies hosted by surviving subhalos as \textit{satellite galaxies}.
    \item \textit{Disrupted subhalos} are halos once identified by {\sc Rockstar} at $z>0$ that no longer exists at $z = 0$ because they deposit the majority of their mass onto the host halo in the interim redshifts. We refer to galaxies hosted by disrupted subhalos as \textit{orphan galaxies}.
\end{itemize}

\subsection{Surviving Subhalos} \label{sec:surv_sub}

We use halo catalogs generated using the {\sc Rockstar} halo-finder \citep{2013ApJ...762..109B} and merger trees generated using the {\sc consistent-trees} merger algorithm throughout this work \citep{2013ApJ...763...18B}. Here we mainly focus on the halo sample at $z=0$. 

We extract all particles and halos around the clusters in CDM and SIDM within a radius of $15 $ \mpch{} of the center. To study the distribution of halos around the clusters, we select halos based on their \vpeak{}, which is the peak value of the the maximum circular velocity within the halo as recorded in the merger-tree catalogs. Peak quantities like \vpeak{} are known to best correlate with luminosities of observed galaxies within subhalos \citep[e.g.,][]{2018ARA&A..56..435W,2021MNRAS.501.1603H}, as current virial masses of subhalos are often affected significantly by tidal stripping in the cluster potential \citep{1983ApJ...264...24M,2019MNRAS.487..653N}.  Galaxies, being more compact, are less likely to be affected by tides and their luminosity traces the original mass of the subhalo before infall \citep{2006ApJ...647..201C}.


In order to select a well resolved population of subhalos, i.e., each having not less than $1500$ particles when \vpeak{} is achieved, a condition of  \vpeak{}$>136.5$ \kms{} is placed on the CDM population of subhalos. The precise value of the cut is based on the abundance matching of SDSS galaxies \citep{2003ApJ...592..819B} and corresponds to all galaxies with r-band magnitude, $M_r<-19.43$ \citep[see e.g., Appendix B in][]{More16}\footnote{The magnitude limit in SDSS corresponds to galaxies with approximately $0.01 L_{\ast}$}. For the SIDM sample, we find that a lower threshold  with \vpeak{}$>130$ \kms{} is appropriate to match the abundance of subhalos (surviving \& disrupted) in CDM. This is partly due to the enhanced disruption in SIDM and also the fact that cored halos in SIDM have smaller maximum circular velocities compared to their CDM counterparts.

\subsection{Disrupted Subhalos} \label{sec:orphans}

Since we use dark matter-only N-body simulations, we use the subhalos as proxies for galaxies in the observed clusters. This is a reasonable assumption in the dark matter paradigm considering all galaxies are expected to reside within their own halos. However, while the overall dynamics of galaxies and subhalos are expected to be similar, there are some subtleties that arise while mapping galaxies to simulated subhalos. Subhalos, being extended objects, are subject to stronger tidal forces and lose mass from their outskirts more easily compared to the galaxies within them. While tidal stripping is the primary mechanism of mass loss for infalling satellites \citep{1983ApJ...264...24M}, enhanced by effects like dynamical friction \citep{RevModPhys.21.383, 1987gady.book.....B} in CDM, subhalos in SIDM can additionally experience evaporation of particles due to self-interactions with both their own particles and host halo particles \citep{2001ApJ...561...61G, Markevitch:2003at}. The time-integrated effect of scattering can be approximated as a net pressure-force given by $\sim \rho v^2  \sigma/m$ \citep{2016MNRAS.461..710D,Kummer170604794} and is often referred to as SIDM ram-pressure. 

While the aforesaid mechanisms constitute modes of physical disruption, artificial disruption can arise due to numerical discreteness effects (e.g., \citealt{VandenBosch171105276,VandenBosch180105427}) and due to the mass resolution threshold inherent to the simulation. In particular, subhalos are not tracked once their mass passes below a resolution threshold, and the rate of mass loss preceding this may be artificially enhanced. Therefore, even if a subhalo ceases to exist in a halo catalog, this does not necessarily imply that the orphan galaxy within it gets disrupted. To alleviate these issues, orphan modeling is often necessary to understand the full distribution of observed galaxies in a given dataset. In this section, we describe briefly our disrupted subhalo tracking and orphan modeling methods.

\subsubsection{Subhalo Tracking Methodology}

To track disrupted subhalos we look up the merger histories of each subhalo generated by the {\sc consistent-trees} algorithm. To find these halos, we track any subhalo that enters the virial radius of its host halo at any point in the simulation and subsequently disrupts. We only study disrupted subhalos with \vpeak{} above the threshold value of 136.5 \kms{} for CDM and 130 \kms{} for SIDM.

In this work, crucially, we use the most bound dark matter particle (MBP) of the disrupted subhalo as a tracer of the location of the associated orphan at $z=0$.  This is a standard choice in CDM simulation analysis because galaxies are expected to be located at the minimum of their halos' potential wells \citep{2004MNRAS.352L...1G,2006MNRAS.366..499D,2011MNRAS.413..101G,2016MNRAS.457.1208H}. This choice also anticipates our weak-lensing studies in Section \ref{sec:sublensing}, which explore the distribution of matter around predicted galaxy locations. Thus, instead of using orbit modeling \citep{2017MNRAS.471.4170T,Nadler180905542,2019MNRAS.488.3143B}, where the orbit of the disrupted subhalo is predicted to infer its $z=0$ position, we use actual particles associated with the disrupted subhalo to trace orphan locations. As our subhalos have $\sim 1000$ particles at peak mass we do not expect that the results will change significantly if we used a different modeling method. We treat our SIDM subhalos in a similar fashion. However, we note that using a single particle as a tracer for disrupted systems in SIDM, makes it susceptible to being scattered out from the minimum of the potential well \citep{2016MNRAS.461..710D} due to interactions with the particles of the host. We test the robustness of MBP as a tracer for a disrupted subhalo in Appendix \ref{conv_test} by looking at the $z=0$ positions of few of the particles located around the MBP at the time when it is selected. 

To isolate the MBP for each of the disrupted subhalos, we look at the snapshot corresponding to the redshift at which its \vpeak{} was attained, following which we find the particle with the minimum mechanical energy $E$ in the reference frame fixed at the center of the subhalo. We model the potential energy, $V$ of the particles using the functional form of that expected for an NFW halo \citep{2008gady.book.....B} for both CDM and SIDM, using the scale radius $r_S$ that has been calculated by {\sc Rockstar} using only the constituent particles of the subhalo:
\begin{equation}
V(r) = - \frac{4 \pi G \rho_0 r_S^3}{r} {\rm ln} \left( 1 + \frac{r}{r_s} \right) 
\end{equation}
We assume that the true potential in a SIDM halo does not deviate much from our `model' NFW potential and hence does not affect our conclusions significantly. We accept only those disrupted subhalos that have >10 particles within $0.25 r_{200}$, with $r_{200}$ representing the virial radius of the subhalo, at the redshift of \vpeak{}.\footnote{In rare cases, subhalos are left out of the catalog because they do not have any bound particles (i.e., particles with total energy $E<0$) according to our method.} Having identified the MBP for each of the disrupted candidates, we find their position in the snapshot at $z=0$.

\subsubsection{Orphan Contribution Model}

When we trace the disrupted subhalos using their MBPs, the mode of their disruption---i.e., whether it is physical or artificial---is not specified by our model. Furthermore, the dominant mode of disruption may differ between the CDM and SIDM models. While artificial disruption due to the mass resolution limit is plausibly the dominant source of disruption in CDM simulations (e.g., \citealt{2021arXiv210301227G}), SIDM subhalos may experience both increased amounts of physical disruption due to ram-pressure stripping as well as numerical effects due to potential biases in halo-finding algorithms \citep{2020ApJ...896..112N}. In light of these uncertainties, \citep{2017MNRAS.469..749P,2021MNRAS.502..621J}, we apply a simple model to associate disrupted subhalos with orphan galaxies, the details of which are as follows:

We allow a fraction of the disrupted subhalos selected using the \vpeak{} threshold to host orphan galaxies within them. Denoting the number of surviving and disrupted subhalos in a given sample as $N_{\rm surv}$ and $N_{\rm dis}$ respectively, we define the orphan galaxy fraction $f_{\rm orp}$ as a function of a free parameter $\eta$ that can be continuously varied between 0 and 1,
\begin{equation} \label{eq:fraceta}
f_{\rm orp}(\eta) = \frac{\eta N_{\rm dis}}{N_{\rm surv} + \eta N_{\rm dis}}
\end{equation}
The free parameter $\eta$ encapsulates our ignorance in assigning orphan galaxies to the disrupted subhalos. It can be also thought of as the probability of a disrupted subhalo hosting an orphan ($\mathcal{P}_{\rm sat,dis}=\eta$); in contrast, we assume that all surviving subhalos host satellites ($\mathcal{P}_{\rm sat,surv}=1$). The extreme values of $\eta=1$ and $\eta=0$ correspond to all and none of the candidates hosting orphans respectively. In detail, $\eta$ depends on the accretion and disruption history of the disrupted progenitor, and it is interesting to explore this dependence in a future work. We emphasize that the interpretation of $\eta$ depends on the mass resolution limit of the simulations. 

\section{Host and Subhalo properties}  \label{sec:subdistrib}

\subsection{Density Profiles of Hosts} \label{sec:host_halo}

\begin{figure}
\includegraphics[scale=0.4,trim={0 0.3cm 0 0.2cm},clip]{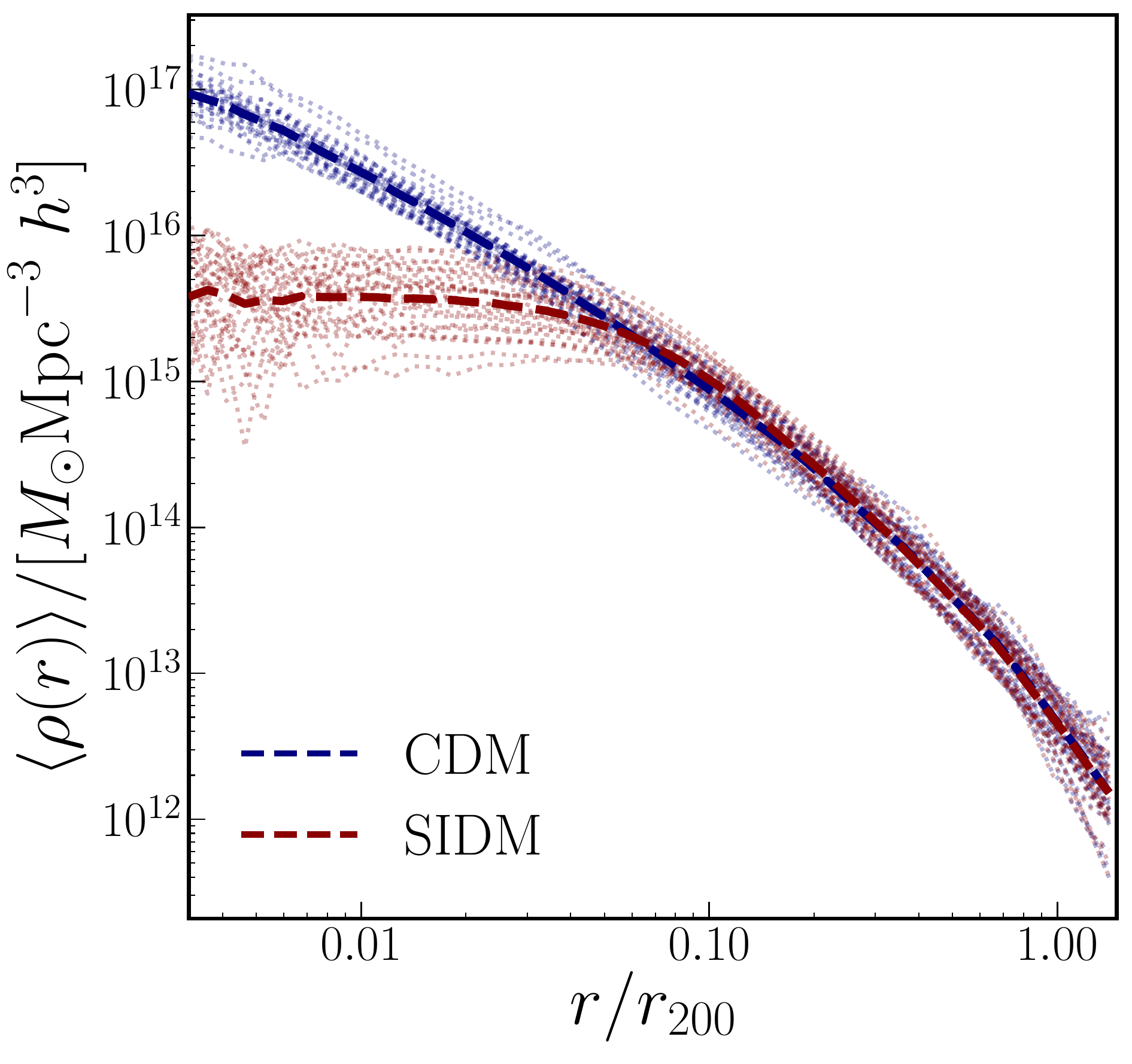}
\caption{The dark matter density profile of the cluster hosts (\textit{faint dotted}) and their mean (\textit{dashed}). Prominent core and cusp features are visible at the centers of the SIDM and CDM hosts, respectively.}
\label{Figure:HostDensity}
\end{figure}

First, we study the distribution of dark matter particles in the cluster halos in zoom-in CDM and SIDM simulations. Fig.\ \ref{Figure:HostDensity} depicts the density profiles of these clusters as a function of radius normalized by the virial radii of the hosts, $r_{200m}$ as determined by {\sc ROCKSTAR}. We confirm on average that the mean profile shows a core within the scale radius for the SIDM clusters. This is expected since the number of scatterings per Hubble time is significantly high for particles within this radius. Furthermore, the density profile in SIDM is higher and steeper than its CDM counterpart right outside the core of the cluster ($r \sim 0.1 r_{200}$). Particles which get scattered to higher energies near the center of the halo end up getting transferred to larger apocentric orbits \citep{2013MNRAS.430...81R}. These contribute to a slight increase in density at such radial distances. We also note that the scatter in the profiles for the SIDM cores is larger than the CDM cusps. 

\subsection{ \vpeak{} distribution of Subhalos} 

\begin{figure}
\centering
\includegraphics[scale=0.4,trim={0 0.3cm 0 0.2cm},clip]{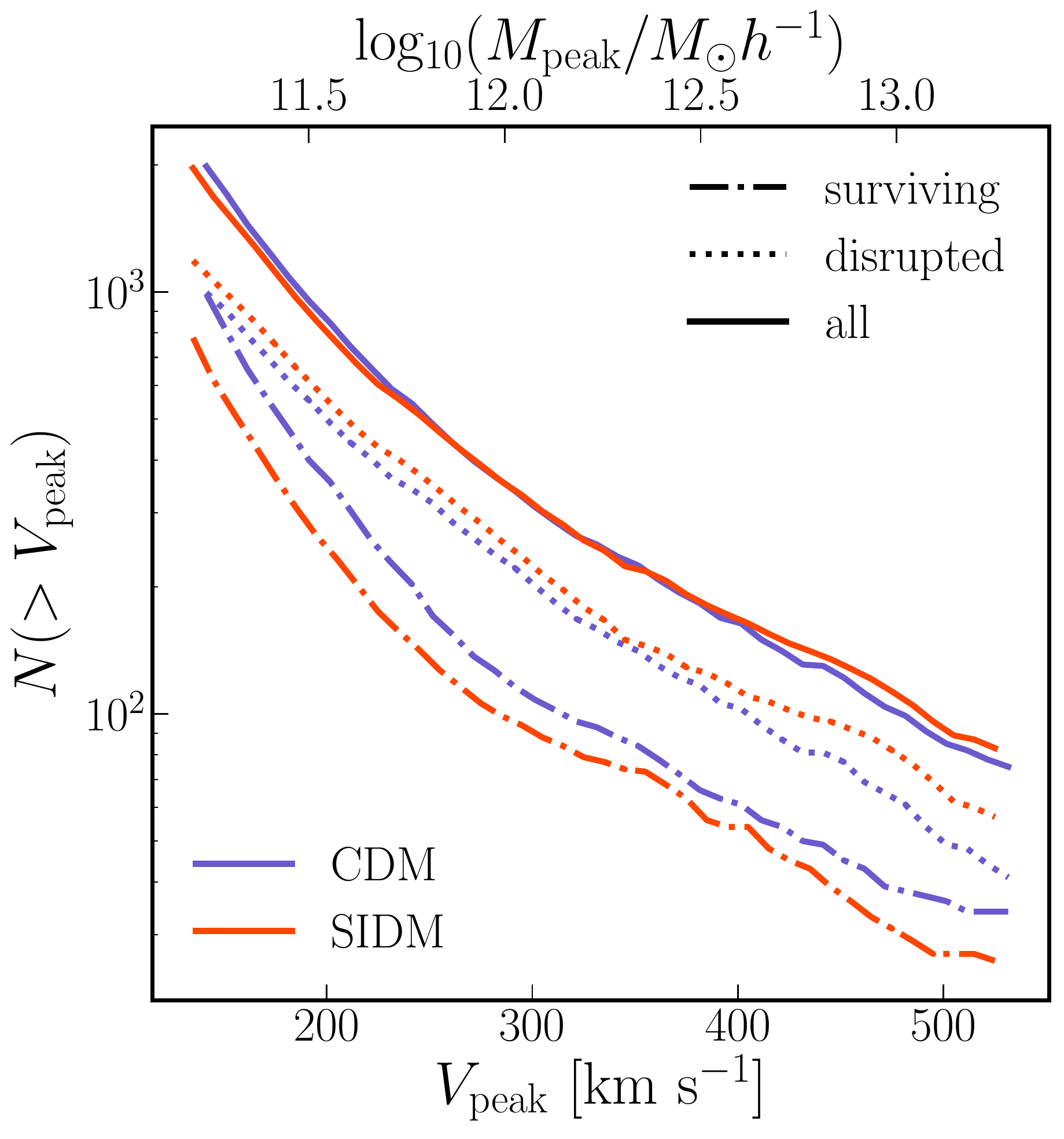}
\caption{Comparison of the cumulative \vpeak{} distribution of all surviving (\textit{dash-dot}), disrupted subhalos (\textit{dotted}) and their combined sum (\textit{solid}). The abundance of the combined population closely match between CDM and SIDM. }
\label{Figure:CCFVpeak}
\end{figure}

Fig.\ \ref{Figure:CCFVpeak} depicts $N(>V_{\rm peak})$, the complementary cumulative distribution of \vpeak{} of the separate and combined populations of surviving and disrupted subhalos in our simulations. The numbers of surviving subhalos are significantly suppressed in SIDM relative to CDM. As a result this gives rise to a larger number of disrupted subhalos in SIDM. When comparing the total population of subhalos we note that \vpeak{} distributions agree well with each other. However, we also observe that the \emph{total} number of low-mass subhalos is slightly suppressed in SIDM, presumably as these subhalos are disrupted before they fall into the massive cluster; this can happen as low mass subhalos can often enter their current hosts as parts of groups, and group environments may have disrupted low-mass subhalos in SIDM even before infall \citep{Nadleretalinprep}. 

\subsection{Radial distribution of Subhalos}

\begin{figure}
\includegraphics[scale=0.4,trim={0 0.3cm 0 0.2cm},clip]{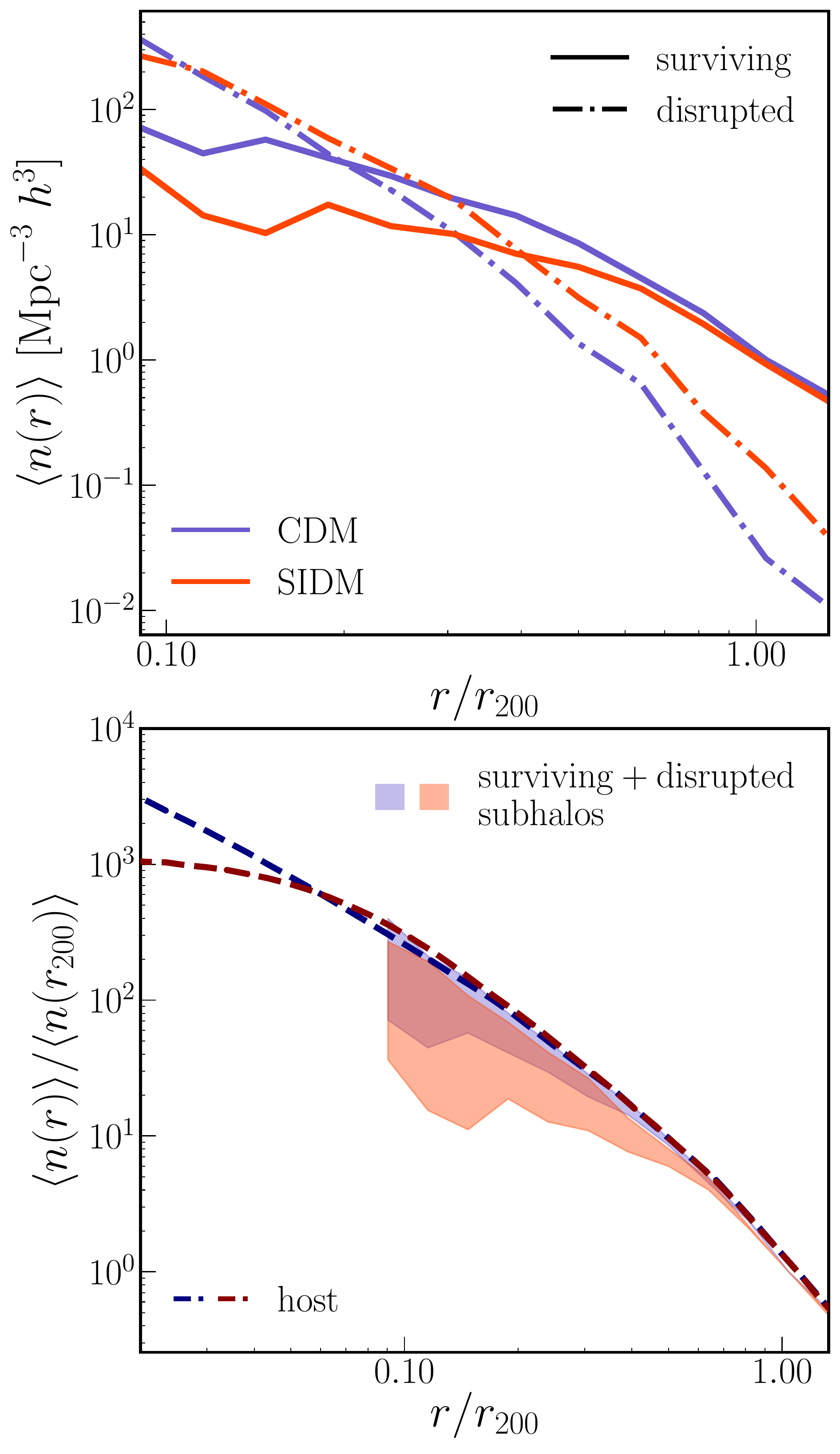}
\caption{\textit{Top}: The stacked radial distribution of CDM (SIDM) subhalos with \vpeak{} > 136.5 (130)~\kms{} with the surviving subhalos $\langle n \rangle_{\rm sat}$ (\textit{thick solid}) and the disrupted subhalos selected using the most bound particle $\langle n \rangle_{\rm orp}$ (\textit{thick dashed}). \textit{Bottom}: The stacked radial distribution of dark matter particles (\textit{dashed}) and satellites (\textit{shaded} regions), including orphan galaxies for orphan fractions $\eta\in [0,1]$. All distributions are plotted as a function of clustercentric radial distances normalized with respect to the cluster virial radius $r_{200}$. }
\label{Figure:SatDensity}
\end{figure}

Here we compare the number density of subhalos in concentric shells around the cluster center in the CDM and SIDM hosts. We evaluate the 3D number density, $n(r)$, in twelve logarithmically spaced radial bins in units of $r_{200}$ of their hosts in the range, $0.08< r/r_{200} < 1.5$. When dealing with the surviving subhalos,  we do not use the subhalo tags from {\sc Rockstar}, instead we compute the number density of all resolved CDM (SIDM) halos above a \vpeak{} threshold of 136.5 \kms{} (130 \kms{}) around the hosts. In Fig.\ \ref{Figure:SatDensity} we show the stacked radial distribution of the surviving subhalos at $z=0$ in the upper panel with the thick solid lines. The number of surviving halos is significantly reduced in the SIDM simulations. This effect persists out to nearly the virial radius of the clusters. 

The distributions of the disrupted subhalos, traced by their MBPs, as a function of clustercentric radius, are shown with the dot-dashed lines in the same figure. The fractional difference between the CDM and SIDM clusters increases as we move outwards. In other words, the drop-off in the number of disrupted subhalos is steeper in CDM compared to SIDM, implying enhanced disruption in SIDM through the halo's interior compared to CDM. 

In the lower panel of Fig.\ \ref{Figure:SatDensity} we compare the subhalo densities to the particle densities in the simulation. We normalize the the number density at a given radius by the number density at $r_{200m}$. The dotted lines correspond to the normalized number density of dark matter particles and the shaded regions correspond to the subhalo distributions. The upper and lower envelopes of the shaded region represent the radial distribution of subhalos in the scenarios when all and none of the disrupted subhalos are taken into account. The shaded region is meant to demonstrate how adding in different fraction of the disrupted subhalos to the surviving population changes the radial distribution. If we assume orphan galaxies populate only a fraction of the disrupted subhalos, the total number density of substructure in any radial bin will be a weighted mean of the contribution from the surviving and disrupted population,
\begin{equation}  \label{eq:nfracorp}
\langle n (r,\eta) \rangle = (1 - f_{\rm orp}(\eta)) \, \langle n (r) \rangle_{\rm surv} + f_{\rm orp}(\eta) \, \langle n (r) \rangle_{\rm dis}
\end{equation}
where $\langle n \rangle_{\rm surv}$ and $\langle n \rangle_{\rm dis}$ are the stacked number density profiles of surviving and disrupted subhalos respectively. The shaded region corresponds to the range $\eta \in \{0,1\}$. 

We observe that the full radial distribution of subhalos, including surviving and disrupted ($\eta=1$, upper envelope), agrees quite well with the dark matter distribution both in CDM and SIDM. This is consistent with the results of \citet{2016MNRAS.457.1208H,2019MNRAS.490.5693B,2021arXiv210301227G}, and here we confirm that it holds for SIDM as well to a large extent. However, we note that the while the CDM subhalo distribution can be as steep as the dark matter distribution (in fact it can even be slightly steeper than dark matter) the SIDM subhalo profile never becomes quite as steep as the dark matter profile within $0.8$ Mpc $h^{-1}$. This can be partly attributed to the fact that the dark matter profile itself becomes steeper in SIDM right outside the core, as dark matter particles are pushed outside from the center region. This phenomena is specific to particles, and does not necessarily effect the profiles of subhalos. A comparison between matter and subhalo distribution can therefore be a probe of the dark matter physics in this regime.

\begin{figure}
\includegraphics[scale=0.4,trim={0 0.3cm 0 0.2cm},clip]{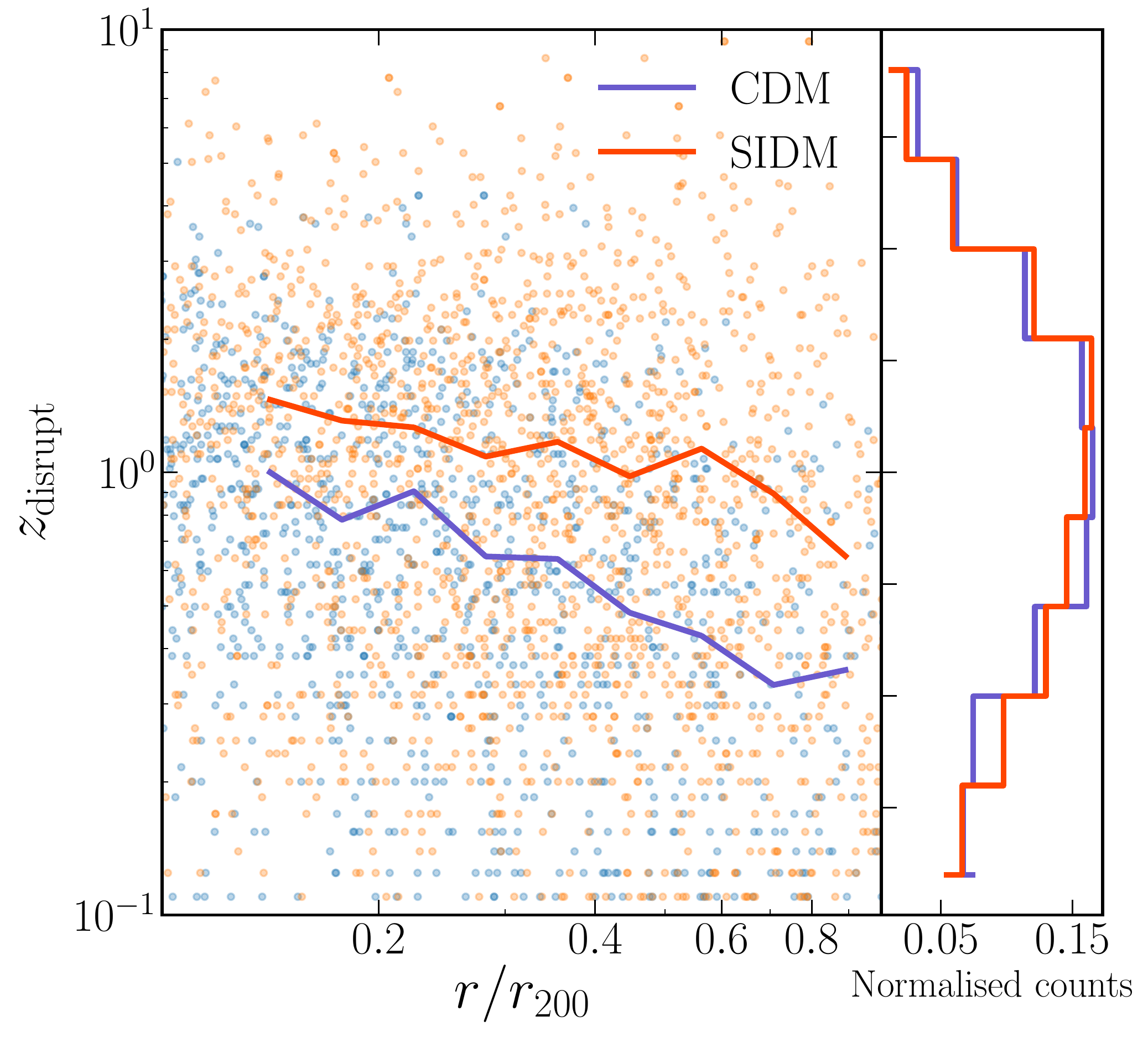}
\caption{The dependence of $z_{\rm disrupt}$, the scale at which the disrupted subhalos were last resolved as halos, on the $z=0$ clustercentric distances of the orphan tracers. Subhalos tend to disrupt earlier in SIDM compared to CDM. In both models, the oldest disrupted subhalos are concentrated near the host center.}
\label{Figure:orp_disrupt}
\end{figure}

In Fig.\ \ref{Figure:orp_disrupt} we show the radial distribution of the redshift of disruption for subhalos in the SIDM and CDM simulations. In general, we find that subhalos tend to disrupt earlier in SIDM compared to CDM.

\section{Lensing around subhalos} \label{sec:sublensing}

Current and future lensing surveys like HSC-SSP \citep{2019PASJ...71..114A}, DES \citep{2018PhRvD..98d3526A} and the LSST \citep{2019ApJ...873..111I} will give us an unprecedented sample of clusters and member galaxies allowing us to measure the detailed mass distribution around subhalos in clusters using satellite galaxy-galaxy weak lensing \citep{2014MNRAS.438.2864L,2015MNRAS.454.3938S}. In this section we measure the matter distribution around surviving subhalos and evaluate the projected excess surface density profile, \esd{}, that is the relevant weak lensing observable. We compare the stacked profiles around surviving subhalos at different projected clustercentric distances, to those around isolated centrals with the same \vpeak{} with the aim of probing the effects of dark matter self-interaction between host and subhalo particles, i.e., at the scale of the cluster velocity dispersion. To avoid the biases incurred in not accounting for the disrupted subhalos \citep{2016MNRAS.457.1208H} we also examine the mass distribution around their tracers, i.e. the most bound particles, and study the stacked profiles as a function of the orphan fraction.

\subsection{Stacked 3D Density Profiles} 

\begin{figure*} 
\centering
\includegraphics[scale=0.33,clip]{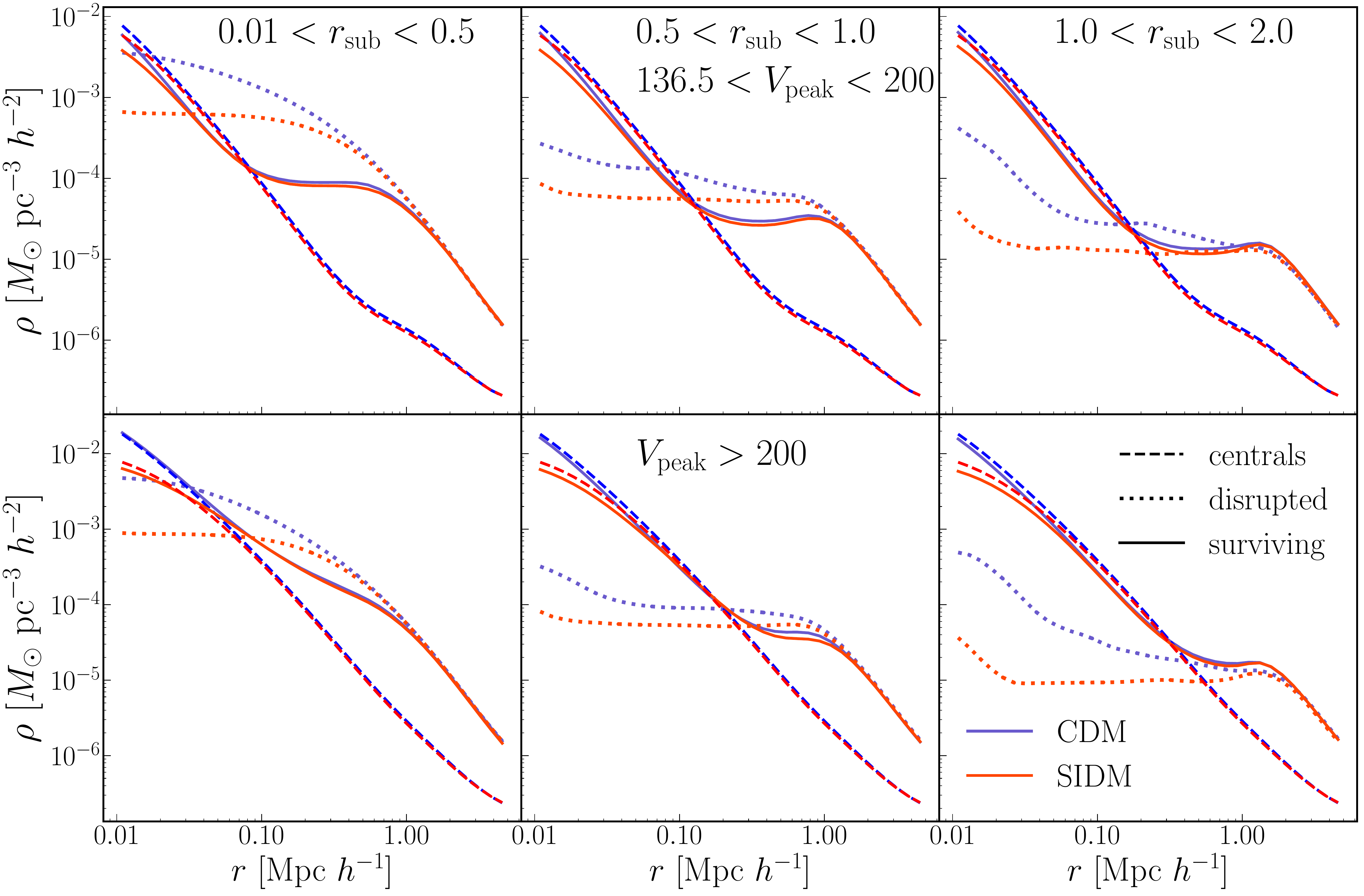}
\caption{The stacked density profiles around surviving subhalos (\textit{solid}) and isolated halos (\textit{dashed}) with \vpeak{}$>136.5$ \kms{} in CDM (\textit{blue}) and SIDM (\textit{red}) as a function of 3D radial distances from the halo centers. Besides this, the stacked density profiles around the disrupted subhalo (\textit{dotted}) around the position of the MBP is shown. While the cores and cusps are present in both SIDM and CDM halos, the surviving subhalos exhibit a contribution from their hosts and tidal stripping throughout their radial extent. There are remnants of cores/cusps of disrupted subhalos is visible in the cluster outskirts. }
\label{Figure:3d_subhalo_profile}
\end{figure*}

Before we study the lensing signal around subhalos, it is instructive to look at the 3D distribution of matter around the subhalos in the simulation directly. In Fig.\ \ref{Figure:3d_subhalo_profile} we plot the stacked density profile of dark matter particles around surviving and disrupted subhalos along with the profiles around isolated halos of the same mass. We study these profiles for subhalos at different clustercentric distances $r_{\rm sub}$ and in two bins of \vpeak. The solid and dashed curves depict the dark matter density around the surviving subhalos and the isolated centrals respectively. In general the core and cusp features are prominent at the centers of the respective SIDM and CDM subhalos, particularly in the massive ones (\vpeak{} $>200$ \kms{}). Two features distinguish the subhalos from the isolated halos: a major contribution from the density peak of the cluster hosts, and a relatively minor suppression in density associated with tidal stripping in the subhalos' inner structure.

The density profile around the MBP tracer for the disrupted subhalos close to the cluster center ($r_{\rm sub} < 0.5$ \mpch{}) is drowned out by the core or cusp features of their respective hosts. On the other hand, the ones beyond 0.5 \mpch{} show signs of feeble remnant of their now disrupted cores or cusps. It is seen that their central densities are suppressed by typically 2 dex relative to the surviving subhalo sample with the same \vpeak{}. The suppression is greater in SIDM as cores are known to be more vulnerable to tidal stripping compared to cusps \citep{2010MNRAS.406.1290P,2021MNRAS.tmp.1267E}.

In the following subsections we study the lensing signal around these objects in detail in finer radial bins.

\begin{figure} \label{Figure:2D_orpsat}
\centering
\includegraphics[scale=0.35,trim={0 0.1cm 0 0},clip]{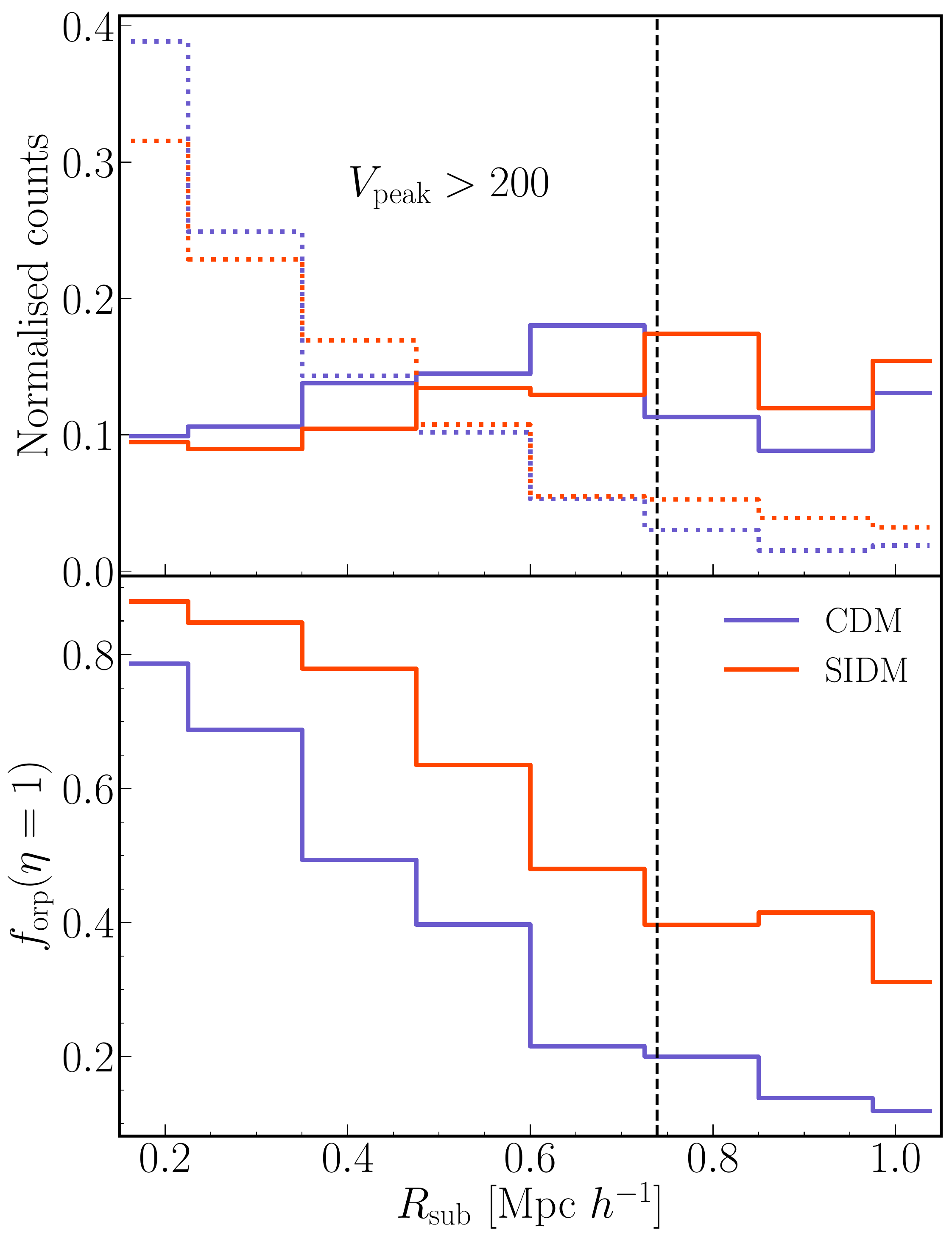} 
\caption{\textit{Top:} The histograms of \rsat{}, the projected clustercentric distances of satellites (\textit{solid}) and orphans (\textit{dotted}). \textit{Bottom:} The maximum possible value of the orphan fraction $f_{\rm orp}$, i.e., with $\eta = 1$, as a function of \rsat{}. The vertical \textit{black dotted} line represents the average value of $0.5r_{200}$ of the cluster hosts. }
\end{figure}

\subsection{Stacked Excess Surface Density Profiles} 

Weak lensing measures the distorted ellipticities of background galaxies behind a lensing source. For any mass distribution the shear field is determined by the excess surface density \esd{} of mass \citep{2005MNRAS.361.1287M,2005astro.ph..9252S}. This is connected to the azimuthally averaged tangential shear field as
\begin{equation} \label{eq:ESDFormula}
\langle \gamma_t (R) \rangle =\frac {\Delta \Sigma (R)}{\Sigma_{\rm crit}}= \frac{\Sigma(<R) - {\Sigma}(R) }{\Sigma_{\rm crit}},
\end{equation}
where $\Sigma(R)$ is the azimuthally averaged projected mass density in a narrow annulus at $R$ and $\Sigma(<R)$ is the average projected mass density integrated within $R$. $\Sigma_{\rm crit}$ is the critical density given by
\begin{equation}
    \Sigma_{\rm crit} = \frac{c^2}{4\pi G(1+z)^2}\frac{D_{\rm s}}{D_{\rm l}D_{\rm ls}}
\end{equation}
which depends on the angular diameter distances to the lens ($D_{\rm l}$), the source galaxies ($D_{\rm s}$), and between the lens and source ($D_{\rm ls}$). 

The value of \esd{} can be measured directly from the dark matter particles in the simulations. We compute the \esd{} profile around the subhalos in our simulations as a function of projected radius $R$ from the center of each subhalo. As the subhalo is embedded in the massive cluster potential there are two separate contributions to the \esd{} profile- one from the enhanced density around the subhalo and the other from the host cluster-mass distribution at the location of the subhalo. 

We measure \esd{} in 20 logarithmically spaced bins between $0.01-5$ \mpch{} centered around the subhalo centers. We count the number of particles in the 2D projected annuli or radius $R$ around every subhalo belonging to the clusters. We assume the $z$ direction as the line of sight direction and compute the projected quantities in the $x-y$ plane. We project over the whole length of the simulation box. For the subhalo population comprising members from all the 30 clusters, we split them into 4 bins according to their projected clustercentric distances, $R_{\rm sub} \in \{ 0.1-0.3, 0.3-0.5, 0.5-0.7, 0.7-0.9 \}$ \mpch{} and study the stacked \esd{} profile in each such bin. 


\begin{figure*} 
\centering
\includegraphics[scale=0.26,clip]{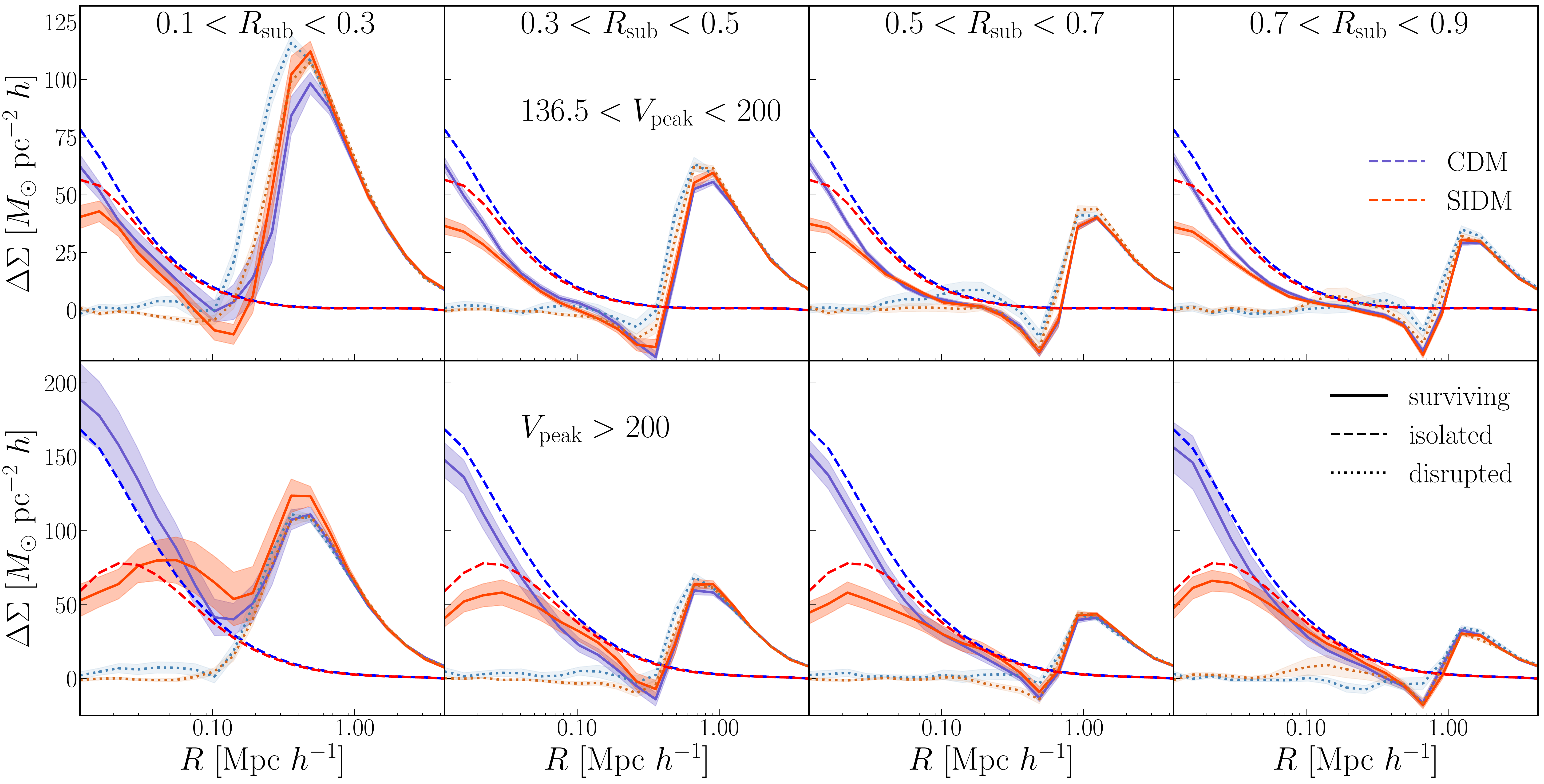}
\caption{The stacked \esd{} profiles for the two bins of \vpeak{} corresponding to the \textit{upper} and \textit{lower} rows and the four columns from the left to right stand for the increasing bins of \rsat{}. The \textit{solid} and \textit{dashed} lines represents the signal from surviving subhalos and isolated centrals respectively, with the shaded band implying the 1$\sigma$ uncertainty. The \textit{faint dotted} curves are the \esd{} profile around the positions of the MBP tracers representing the disrupted subhalos.}
\label{Figure:esd}
\end{figure*}

We compare the stacked \esd{} profiles of subhalos with that of centrals or isolated centrals with the same \vpeak{} threshold. These centrals were selected using the condition that they are not within 5 \mpch{} of any host with mass greater than $10^{13} M_{\odot}$. In a given zoom-in box, the number of centrals was found to be typically an order of magnitude larger compared to the satellites in the cluster.

In Fig.\ \ref{Figure:esd} we show the stacked ESD profiles around the surviving subhalos (solid lines) and isolated centrals (dashed lines) in our simulations. The top and bottom panels show the ESD profiles for two different bins of \vpeak{}. The differences between isolated centrals and subhalos as observed in the 3D density profiles are reflected here. We note that the small radii ($R \lesssim 0.2$ \mpch{}) are dominated by the (sub)halo's own overdensity and at the location of the host center ($R \sim$ \rsat{}), \esd{} which essentially traces the slope of the density profile, changes sign. The overall amplitude of the density within surviving subhalos is suppressed compared to the isolated halos of the same \vpeak{} due to stripping of mass. This effect is significantly more severe for the lower \vpeak{} subhalos, both in CDM and SIDM.

The stripping of CDM subhalos throughout their extent is surprising; tides are expected to strip material from their outskirts, but we observe a depletion throughout subhalos' cusps. A possible explanation for this effect can be that some of the particles in the outskirts have radial or plunging orbits within the subhalo, and they get stripped when they are near the outskirts of their orbits. Moreover, subhalos can become significantly aspherical inside a cluster, which my invaldiate assumptions about their cusps not being disrupted in a spherical potential.\footnote{A peculiar feature is noticed in the innermost bin of \rsat{} for \vpeak{}$ > 200$ \kms{} that has CDM satellites being more dense than CDM centrals, this is due to the cluster particles themselves elevating the number density of particles near the center}.

The stacked \esd{} profiles of the disrupted subhalos (dotted lines) are very flat close to their centers because they have lost most of their mass and the projection makes the profiles appear shallower than in 3D. The net effect of adding these disrupted subhalos to the stacked \esd{} profile of the surviving subhalos, is to suppress the average density profiles inside their virial radius. As both dark matter interactions and disruption lead to a suppression of the density profile around a galaxy, it is important to consistently account for the orphan contribution when making inferences about dark matter microphysics. 

The sample of satellite galaxies that are observed in clusters may correspond in part to ``orphan" galaxies in our simulations, i.e. they may exist in disrupted subhalos. Using the formalism of Section \ref{sec:orphans} we can identify the orphan candidates, and measure the \esd{} profile around their MBP tracers. Because we do not know what fraction of the observed satellites are orphans \emph{a priori}, we allow the fraction to vary before adding their contribution to the stacked subhalo profiles. In order to obtain a meaningful observable we appropriately combine them as a weighted average, 
\begin{equation}  \label{eq:ESDfracorph}
\Delta \Sigma(R,\eta) = \left(1 - f_{\rm orp}(\eta)\right) \, \Delta \Sigma_{\rm sat}(R) + f_{\rm orp}(\eta) \, \Delta \Sigma_{\rm orp}(R),
\end{equation}
where $f_{\rm orp}$ is calculated using Eq. \ref{eq:fraceta}. For context, $\eta=1$ corresponds to the case where all disrupted subhalos host orphans and $\eta=0$ corresponds to the case where none of them do. Note that the fraction of disrupted subhalos itself varies as a function of the projected cluster-centric radius $R_{\rm sub}$ as can be seen in the bottom panel of Fig.\ \ref{Figure:2D_orpsat}. 

In the following section, we explore the degeneracies due to orphan modelling and self-interactions on the \esd{} profiles, and study how well we can distinguish between them two by constructing mock lensing observables for an HSC like data set around satellite galaxies.


\subsection{Mock lensing profiles for satellite galaxies}

\begin{figure*}
\centering
\includegraphics[scale=0.27,clip]{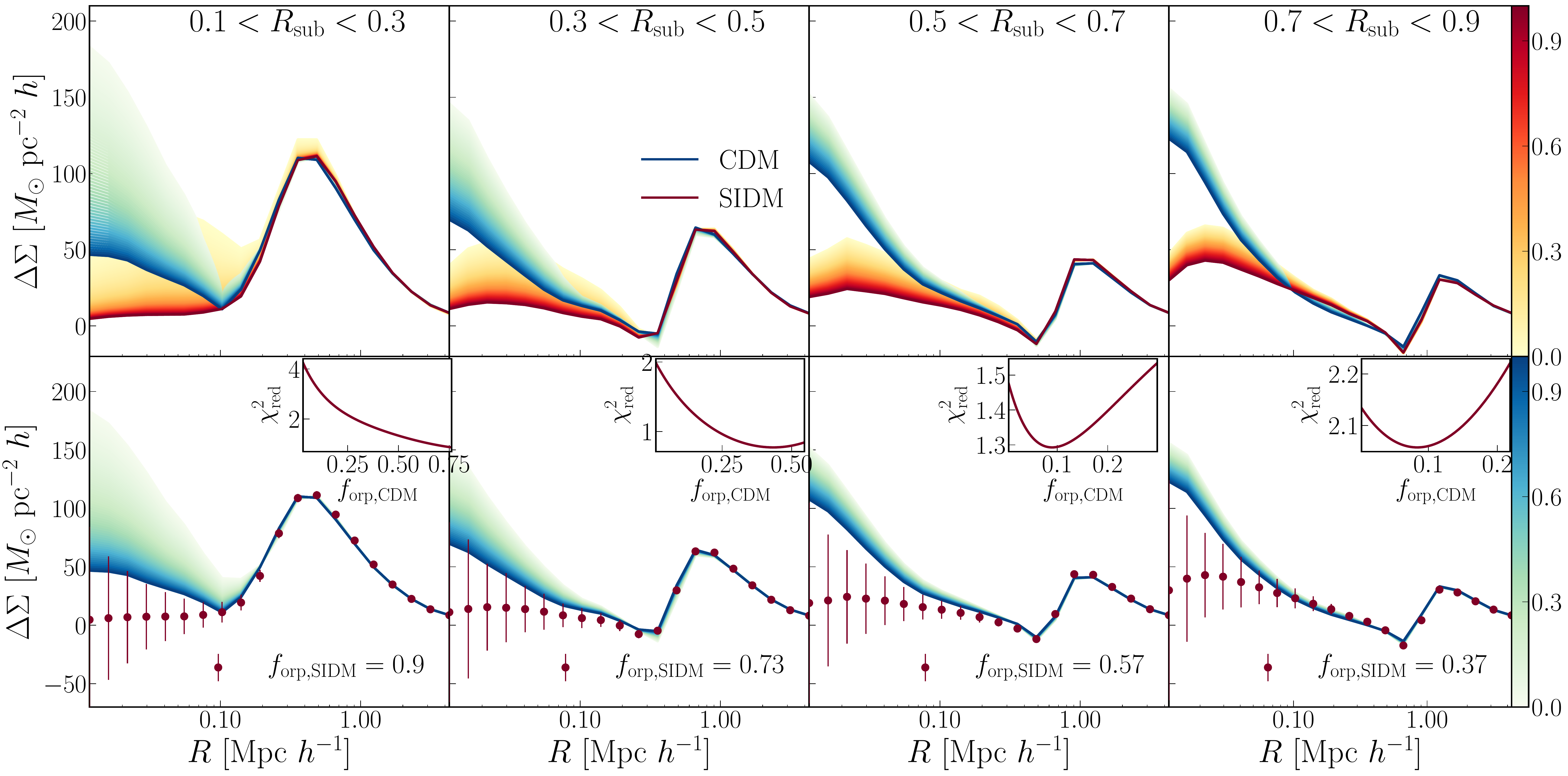}
\caption{The stacked \esd{} profiles for satellites+orphans with \vpeak{} $>200$ \kms{}, created using Eq. \ref{eq:ESDfracorph} at four different distances from the cluster center (in Mpc $h^{-1}$). The \textit{upper} panel depicts the effect of varying $\eta$ for the SIDM and CDM models (colorbar on extreme right), with a darker shades implying a larger orphan fractions. The \textit{bottom} panel shows a mock measurement created from SIDM satellites and orphans with $\eta=1$. The errorbars are derived from the data covariance matrices of the weak lensing shear measured in the HSC-SSP survey \citep{KumarMoreinprep}. The $\chi^2_{\rm red}$ from fitting the mock profiles with the CDM model profiles is shown in the inset plots in the \textit{upper corners}.}
\label{Figure:esd_mock}
\end{figure*}

In the upper row of Fig.\ \ref{Figure:esd_mock} we show the stacked \esd{} profiles for subhalos with \vpeak{}$>195$ \kms{} in SIDM and \vpeak{}$>200$ \kms{} in CDM simulations when different fractions of disrupted subhalos are populated with orphans. From light to dark, the curves correspond to assigning successively higher fraction of orphans $f_{\rm orp}$. The four panels correspond to different bins of projected distances of the subhalo lens from the cluster center, \rsat{}. We note that the disparity between the \esd{} profiles of SIDM and CDM subhalos decrease with their proximity to the cluster center. This is because when $f_{\rm orp}$ is sufficiently high, the cusp of the stacked CDM profiles is damped out and resembles cored like the SIDM profiles, confounding inferences about dark matter interactions. 


In the lower panel of Fig.\ \ref{Figure:esd_mock} we demonstrate a typical scenario that we will encounter while measuring the stacked weak lensing signal in cluster satellites in a universe with SIDM. The red points correspond to mock measurement of the stacked lensing profile in the SIDM simulations when $100\%$ of orphans are assumed to have galaxies, i.e. $\eta=1$; to this profile we add the error bars from the HSC survey. The errors are determined from a cross-correlation of the positions of satellite galaxies in SDSS redMaPPer clusters \citep{2016ApJS..224....1R} with the shear obtained from the first year shape catalog release of the HSC-SSP \citep{2018MNRAS.481.3170M}. These errors shown here reflect the diagonal component of the shape noise covariance $\mathbf{C}_{\rm HSC}$ which has been calculated by 320 different realizations of randomly rotated shapes of HSC galaxies around redMaPPer satellite galaxies \citep{KumarMoreinprep}. We also overlay the CDM profiles with different orphan fractions to demonstrate that given the current error bars, a typical scenario in an SIDM universe can be degenerate with a CDM model that has a high orphan fraction, particularly for satellites near the cluster center. 

With the aim of studying the joint effects of self-interactions and subhalo disruption on the subhalo \esd{} profiles, we conduct a mock observation of an SIDM universe and try to fit it with a CDM model. The fraction of orphans in both the mock and model samples are varied and the goodness-of-fit is checked based on a $\chi^2$ measure. We study the stacked \esd{} profiles for each of the 4 bins of \rsat{}.  We construct the set of mock observations $\boldsymbol\Sigma_{\rm SIDM}$ from the SIDM simulations by varying the orphan fraction through 100 bins of $\eta_{\rm SIDM} \in \{0,1\}$ using Eq. \ref{eq:fraceta}. For our model space, we use the CDM simulated \esd{} profiles $\boldsymbol\Sigma_{\rm CDM}$ for the same range of the parameter $\eta_{\rm CDM}$. For each observation in a given bin of \rsat{} we compute the $\chi^2$ with the model CDM profiles using the weak lensing covariance $\mathbf{C}_{\rm HSC}$. Therefore the $\chi^2$ at each point in our 2D space of mock and model orphan fractions is given by
\begin{equation}
\chi^2_{ij}= (\boldsymbol\Delta \boldsymbol\Sigma^{i}_{\rm CDM} - \boldsymbol\Delta \boldsymbol\Sigma^{j}_{\rm SIDM})^{\mathbf{\intercal}} \mathbf{C}^{-1}_{\rm HSC} (\boldsymbol\Delta \boldsymbol\Sigma^{i}_{\rm CDM} - \boldsymbol\Delta \boldsymbol\Sigma^{j}_{\rm SIDM}),
\end{equation}
where $i$ and $j$ iterate over the bins of $\eta_{\rm CDM}$ and $\eta_{\rm SIDM}$. The number of degrees of freedom in our model is 19, with 20 radial bins around each lens and the free parameter, $\eta_{\rm CDM}$.

\begin{figure*} 

\centering
\includegraphics[scale=0.265,clip]{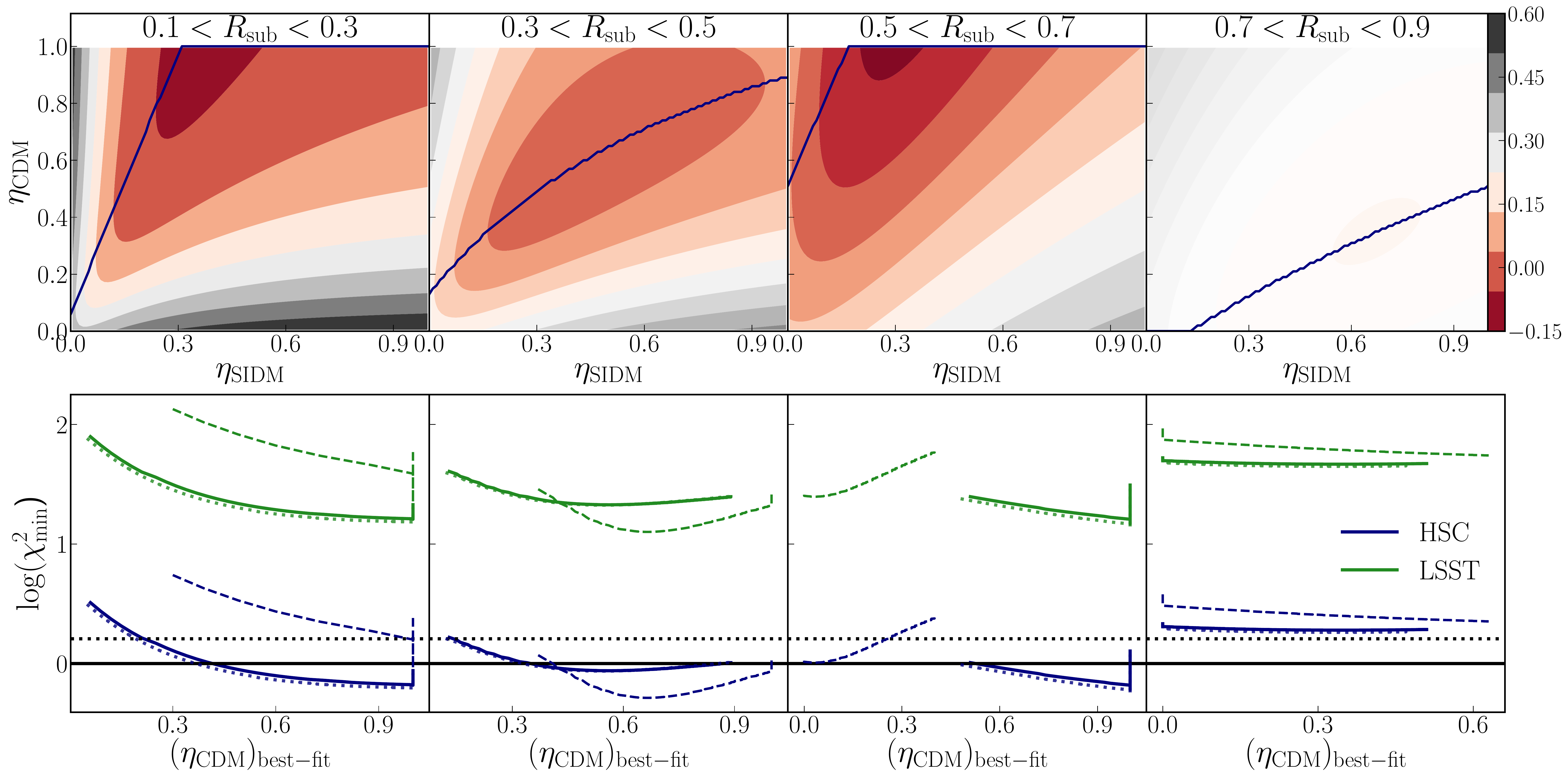}
\caption{\textit{Upper row:} The logarithmic reduced $\chi^2$ obtained by fitting the mock $\Delta \Sigma$ profiles around SIDM subhalos using HSC error bars, with CDM profiles by varying the orphan contribution, $\eta$. The x-axis shows the different values of $\eta_{\rm SIDM}$ used to create the observable, the y-axis shows the model $\eta_{\rm CDM}$. The loci of minima for each $\eta_{\rm SIDM}$ is plotted as a \textit{dark blue} line. The interplay of the effects of inclusion of orphans and self-interactions makes it difficult to disentangle the two for \rsat{} $<0.7$ \mpch{}. \textit{Lower row:} The values of the minimum $\chi^2/{\rm d.o.f.}$ are plotted against the position of the minima, $\eta_{\rm CDM}$. The \textit{horizontal black solid} line represents $\chi^2/{\rm d.o.f.}=1$ and the \textit{horizontal black dotted} line is the $2\sigma$ upper limit for this $\chi^2/{\rm d.o.f.}$ distribution. A larger value of the minimum $\chi^2/{\rm d.o.f.}$, implies a poor fit of the CDM model to the SIDM mock observation, thereby represents a possible way to constrain a SIDM model. The \textit{dotted} lines shows the result of incorporating a stellar mass contribution to the \esd{} profile and the \textit{dashed} lines reflect the effect of miscentering the host. The \textit{dark green} lines shows the expected constraining power of the LSST survey.}
\label{Figure:esd_chi2}
\end{figure*}

We show the 2D distribution of ${\rm log}(\chi^2/{\rm d.o.f})$ in Fig.\ \ref{Figure:esd_chi2}. The $x$-axis and the $y$-axis are $\eta_{\rm SIDM}$ and $\eta_{\rm CDM}$ respectively. We note that in the innermost regions around the cluster, low values of $\eta_{\rm SIDM}$ in an SIDM universe will be inferred as a high $\eta_{\rm CDM}$ in CDM. This is because the density deficit arising due to the cored nature of the SIDM satellites can also be compensated by CDM satellites with a larger contribution of orphans. But as we move to the outer most bins there are no good fits to the assumed model for the observed set of curves. In the outer regions (\rsat{} > 0.7 \mpch{}), an orphan fraction of $>0.2$ in the CDM model would not be permissible to explain the data in an SIDM cosmology, because we find that the fraction of CDM orphans can be at most 0.2 (see Fig. \ref{Figure:2D_orpsat}). Therefore a way to distinguish between SIDM and CDM in a HSC-like survey will be to observe the lensing profiles in the cluster outskirts. 

For each value of $\eta_{\rm SIDM}$ the minimum of the reduced chi-square $\chi_{\rm min}^2$ and with it the $\eta_{\rm CDM}$ at which the minima is obtained are plotted against each other as the solid navy blue line in the lower row of Fig.\ \ref{Figure:esd_chi2}. The solid and dotted black lines represent $\chi^2/\rm{d.o.f}=1$ and the $95 \%$ confidence interval of a $\chi^2$ distribution with $\rm{d.o.f}=19$ respectively. If all of the minima, $\chi_{\rm min}^2$ falls outside this interval, the probability of the \esd{} profile arising from CDM rather than SIDM substructure can be rejected at $>95\%$ confidence independent of the underlying orphan fraction. For bins of \rsat{} other than the outermost one, the SIDM and CDM \esd{} profiles tend to give $\chi_{\rm min}^2$ that fall within the interval. This implies that it is very difficult to constrain $\sigma/m$ using the weak lensing signal from satellites at projected distances $< 0.7$ \mpch{}. However, in the outermost bin, the abundance of orphans decrease enough for their effect on \esd{} to be insufficient in compensating for the reduced density due to self-interactions. As a result $\chi_{\rm min}^2$ remains outside the interval for the full range of the best-fit $\eta_{\rm CDM}$ which implies that this can provide a possible way to place an upper-limit on $\sigma/m$.

\subsection{Projections for the LSST Survey}

We also conduct the same experiment by estimating the lensing covariance for a future survey like LSST. Assuming that the magnitude of the covariance matrix for galaxy-galaxy tangential shear is inversely proportional to the survey coverage area, $\Omega$, and the effective number density of background galaxies, $n_{\rm eff}$ \citep{2015ApJ...799..188S}, $\mathbf{C}_{\rm HSC}$ is scaled by the appropriate ratios such that
\begin{equation}
\mathbf{C}_{\rm LSST} \simeq \frac{n_{\rm eff,HSC}}{n_{\rm eff,LSST}} \frac{\Omega_{\rm HSC}}{\Omega_{\rm LSST}} \mathbf{C}_{\rm HSC}.
\end{equation}
Here, we have used $n_{\rm eff,HSC}=21.6$ arcmin$^{-2}$ \citep{2018PASJ...70S..25M} and $n_{\rm eff,LSST}=37$ arcmin$^{-2}$ \citep{2013MNRAS.434.2121C}. For the sky-coverages of LSST and HSC we use $\Omega_{\rm LSST} \simeq 20000$.   The results are summarized in the bottom panel of Fig.\ \ref{Figure:esd_mock} with the dark green solid line. Based on our estimation we note that the LSST will shrink the error bars significantly making it easy to rule out the effects of subhalo disruption in an SIDM universe when we measure the \esd{} profiles around satellites.


In summary this demonstrates the challenges that we will face in using the weak lensing observable \esd{} to infer the effects of self-interactions. Irrespective of the nature of dark matter, the orphan fraction should be allowed to be a free parameter when \esd{} is measured \citep{KumarMoreinprep}. Ideally if one can constrain the orphan fraction as a function of radius, independently using the radial distribution of subhalos (Fig.\ \ref{Figure:SatDensity}) and then use it as a prior for the weak lensing analysis, the nature of dark matter interactions can be inferred. Alternatively, the weak lensing profiles around the satellites in the cluster outskirts may be used because in these regions the disrupted remnants are less abundant. 

In the final analysis section we discuss some possible systematics that need to be considered for a realistic model.

\subsection{Stellar Mass Contribution and Miscentering}

Here we explore how the contribution of the baryonic mass in the satellite galaxies affects the weak lensing profiles and therefore our inferences. Stellar masses $M_{\star}$ are assigned to the subhalos using a $M_{\star}$-\vpeak{} relation from \citet{2018MNRAS.477..359C}, from which we derive \esd{}$_{\star}$ profile by assuming the galaxy to be a centrally located point mass. The details of our method are described in Appendix \ref{sec:baryons}. Furthermore, we also test the effects of cluster miscentering \citep{2008JCAP...08..006M} on the subhalo profiles. Miscentering may be significant source of systematic uncertainty as it will cause subhalo distances from the cluster center to be mislabelled. Our method of introducing miscentering in the simulated \esd{} profiles is described in Appendix \ref{sec:miscent}. 

The impact of including stellar mass and miscentering on the $\chi^2_{\rm min}$ estimate is depicted by the dotted and dashed lines respectively in the lower panel of Fig.\ \ref{Figure:esd_chi2}.
Although the addition of a stellar component may appear to drastically change the dark matter only \esd{} profile compared to the feeble effect of miscentering (see Fig.\ \ref{Figure:esd_syst}), these two have opposite effects on the $\chi^2$ estimate. This is because the covariance for the inner radial bins ($R < 0.01$ \mpch{}) is much larger compared to the radial bins at $R \sim 1$ \mpch{}, e.g., the errorbars reflect this in Fig.\ \ref{Figure:esd_mock}. Therefore, systematics at large scale like miscentering contribute to the $\chi^2$ more than baryonic effects at small-scales. Nonetheless, Fig.\ \ref{Figure:esd_chi2} show that, satellites with \rsat{} >0.7 \mpch{} could be still used to probe SIDM as miscentering leads to poorer rather than better fits to CDM. 

Our miscentering estimates are derived using the miscentering fractions for the RedMaPPer cluster finder \citep{2017MNRAS.469.4899M}. Optical cluster finders like RedMaPPer, that assign centers to bright central galaxies have larger fractions of miscentered clusters than X-Ray or SZ selected clusters \citep{2019MNRAS.487.2578Z}, therefore the effects of miscentering can be mitigated by using alternative cluster finders. 


\section{Discussion $\&$ Outlook}

In this section we highlight key takeaways from our analysis and some caveats that must be accounted for to make robust inferences for dark matter physics based on comparisons of cluster satellites to cosmological simulations.

\subsection{Artificial Disruption and Satellite Abundances}

While \citet{VandenBosch171105276,VandenBosch180105427,2021MNRAS.tmp.1267E} discuss forms of numerical disruptions inherent in CDM simulations, an equivalent study with the same detailed analysis for SIDM is yet to occur. By focusing on satellites hosted by relatively well-resolved subhalos (for example, we only studied SIDM subhalos with \vpeak{} > 195 \kms{} for the lensing analysis), we have mitigated the impact of artificial disruption; however, this effect may become severe for less massive substructure in SIDM due to evaporation \citep{2001ApJ...561...61G}.

In \citet{2013MNRAS.430...81R} the subhalo counts for cluster hosts are suppressed in a $\sigma/m =1$ cm$^2$~g$^{-1}$ realization of SIDM by a few percent relative to the CDM equivalent, especially in the inner region of the halo ($r<0.5r_{vir}$). Our results for surviving subhalos seem consistent with this result, however when disrupted subhalos are taken into account, both SIDM and CDM subhalos are found to be equally abundant within clusters. Therefore, SIDM is comparable with CDM in being able to explain the abundance of massive substructure in clusters \citep{Moore9907411,2017MNRAS.468.1962N}. This is expected if all halos that form in CDM also form in SIDM, given that the primordial matter power spectrum is the same for both models down to the mass scales we are observing in this paper \citep{2016MNRAS.460.1399V,2018PhLB..783...76H}. However, there may still be small differences in the SIDM subhalo abundance in clusters from disruption within larger groups before infall into a larger host halo \citep{Nadleretalinprep}. 

To use the abundance of satellites in clusters for constraining models of dark matter physics, whether using studies of the spatial distribution of bright satellites \citep{Budzynski:2012yd, Shin:2021odx} or lensing mass maps \citep{2017MNRAS.468.1962N}, it is therefore imperative to understand how galaxies populate their subhalos and how subhalo disruption is related to the disruption of the galaxy within them. While central dark matter cusps are never completely disrupted in CDM \citep{2021MNRAS.tmp.1267E}, the existence of cores in SIDM subhalos makes these systems more susceptible to disruption \citep{2010MNRAS.406.1290P}. In addition, the central cores may lead to more extended satellite galaxies with shorter tidal disruption timescales compared to CDM. A detailed study of this effect is required to interpret observed satellite populations in the context of SIDM.

\subsection{Baryonic Contribution and Galaxy-halo Connection}

In principle, to get a complete picture of the evolution of satellite galaxies in clusters, we need to robustly account for their baryonic component and its influence on dark matter \citep[e.g.,][]{2000ApJ...542..535G,2002MNRAS.333..177B,2014ApJ...786...87B,2015MNRAS.451.1247S,2016MNRAS.457.1931S,Kim1812121,Nadler171204467}. In this paper we model the baryonic contribution to the \esd{} profiles in an ad hoc manner (see Appendix \ref{sec:baryons}) and therefore do not account for co-evolution of the galaxy's dark matter halo and its baryonic component.  

There are two primary ways in which baryons affect their halos. Firstly, adiabatic contraction of dark matter orbits in the presence of a central galaxy can make the centers of halos appear cuspy \citep{Gnedin:2004cx}; it has been shown that this effect is enhanced in SIDM halos making them appear as cuspy as CDM halos, if not more \citep{Despali:2018zpw, 2021arXiv210212480S, Kaplinghat:2013xca}. Secondly, feedback processes from supernovae \citep{2012MNRAS.421.3464P} and AGN \citep{2017MNRAS.472.2153P} remove dark matter particles from the center creating a core. Both these competing effects can, in principle, make the density profiles of subhalos in SIDM appear degenerate with a scenario that has both CDM and baryons, particularly within $0.01$ \mpch. However, this is unlikely to change our inferences from the stacked \esd{} profiles as most of our signal comes from the enhanced stripping of SIDM subhalos due to interaction between host and subhalo particles, which impacts the profiles throughout the subhalos entire extent.

Several studies have been conducted on the evolution of Milky Way satellites in the presence of SIDM \citep{Fry150100497,Robles170607514,2019MNRAS.490.2117R,Fitts181111791,2018MNRAS.479..359S,2018ApJ...853..109E, 2016MNRAS.461..710D}. On cluster mass scales \cite{Robertson:2016xjh} and \cite{Robertson:2020pxj} have studied the evolution of cluster in the presence of baryons, but have focussed mostly on host profiles. We note that the presence of a massive central galaxy in the cluster can also affect the survivability of satellites that are on highly radial orbits-- this effect is more severe in disk centrals due to its axis-symmetric potential. On Milky Way scales, it has been shown that the disruption of satellites due to the central disk can be significant in shaping the abundance of satellites and their radial distribution. On cluster scales, however, this effect is less severe considering cluster centrals are mostly elliptical and generally the bright satellite galaxies have been on fewer orbits within a cluster compared to satellites of the Milky Way. 

Finally, it has recently been pointed out that the timescale for gravothermal core-collapse can be shortened for satellite galaxies at large interaction cross-sections ($\gtrsim 10$ cm$^2/$g), due to tidal stripping of their outer profiles \citep{Nishikawa190100499, Kaplinghat190404939,2021MNRAS.503..920C}. This can further accelerate in the presence of baryons that can aid adiabatic contraction and generally the satellite survival probability \citep{Haggar:2021sqo}. However, we note that the cross-section at the satellite mass-scales investigated in this paper are unlikely to be much higher than a few cm$^2$g$^{-1}$, making these systems unlikely candidates for core-collapse \citep{Kaplinghat:2015aga}. Moreover, because clusters are relatively young objects, a large fraction of satellites are unlikely to have been within the cluster potential for longer than a few orbital timescales, making them less susceptible to enter the collapse phase.


Given these subtleties and open questions, it is important to explore the galaxy-halo connection in an SIDM universe to account for the interplay of baryonic and dark matter physics in these models. Our work is a step towards disentangling some of the degeneracies in the domain of dark matter-only simulations. Our findings imply that while the radial distribution of subhalos, including both surviving and disrupted systems can be quite similar in CDM and SIDM, the inner structure of these objects can be significantly different. Furthermore, these differences can significantly affect the stellar properties of galaxies that live within them, and vice versa \citep{2016MNRAS.461..710D}.

\subsection{Predictive Power of Weak Lensing}

Measurements of the weak lensing profile around satellite galaxies have previously been used to infer the dark matter distribution around them \citep{2015MNRAS.454.3938S,2018MNRAS.478.1244S}. The shape of the weak lensing profile around the satellites is a sensitive probe of the velocity dependent interaction cross-section as it depends both on the cross-section at the low mass subhalo scale and the cluster mass scale. However, we show that inferences about dark matter can be complicated by degeneracies between the galaxy occupation in disrupted subhalos and the SIDM model. The contribution to the total weak lensing profile around satellite galaxies that comes from disrupted subhalos, which have dramatically stripped dark matter profiles needs to be accounted for to accurately infer the effect of particle interactions. 

To forward model the weak lensing signal around satellite galaxies from a given simulation, one can first attempt to constrain the contribution from disrupted subhalos ( orphan galaxies  ) by measuring the number of observed galaxies as a function of radius. This will allow us to use the orphan fraction as a prior to estimate differences from CDM subhalo profiles. Alternatively, the weak lensing signal can be measured around satellite galaxies near the outskirts of the cluster (\rsat{} $> 0.5r_{200}$) where there are fewer disrupted systems (Fig \ref{Figure:2D_orpsat}). However, we expect the exact value of the inner radial cut-off region to depend on sample of galaxies that we observe.

In this work, we have selected subhalo samples with \vpeak{} thresholds to match the total number of subhalos in CDM and SIDM. This choice was made envisioning a measurement where the satellite galaxy sample is selected based on a luminosity threshold in observations and its corresponding subhalo sample is selected based on abundance matching. We set the abundance using subhalos from CDM with a \vpeak{} threshold that is matched to the abundance of typical galaxies used for cluster studies in galaxy surveys. However, as noted previously, we find that the same abundance is obtained in SIDM at a smaller \vpeak{} (see Sec. \ref{sec:sims}), implying that the SMHM relation in SIDM can potentially be different. Alternatively, we can also obtain the appropriate \vpeak{} selection by directly measuring weak lensing profiles around \textit{isolated} galaxies at a given luminosity and use it for our simulation comparisons between CDM and SIDM. We find that using the latter method, as expected, the \vpeak{} inferred from the stacked \esd{} profiles of SIDM, isolated subhalos is smaller than CDM (due to coring of the profiles). When subhalos are selected using the \vpeak{} inferred from isolated profiles, we find that the degeneracies between SIDM and CDM become generally more severe. However, we find that our overall inferences do not change significantly, i.e. subhalos in the cluster outskirts are still the most promising probes of dark matter and observations from the LSST survey should capture the subtle changes throughout the subhalo profiles and help constrain deviations from CDM.

\section{Result \& Conclusion}

We run 30 N-body, zoom-in simulations of cluster-mass (>10$^{14}$ \mpch{}) dark matter halos with a velocity dependent SIDM cross-section to conduct a detailed comparison of the distribution and properties of the massive subhalo population with peak velocity \vpeak{} $\gtrsim 130$ \kms{}. Throughout, we have aimed to consistently account for the population of disrupted subhalos in the dark matter simulations and understand their impact on the statistics of various observables that can potentially help understand the nature of dark matter. In particular, we have focused on the subhalo radial distribution and the weak-lensing profile around subhalos in observed galaxy clusters. The principle findings of our work are as follows,

\begin{enumerate}[--]

    \item The potential remnants of disrupted subhalo and satellite galaxies must be accounted for in order to generate robust predictions for subhalo and satellite populations from SIDM simulations.
    
    \item The \vpeak{} and radial distributions of the combined sample of disrupted and surviving subhalos in the CDM and SIDM scenarios agree with each other reasonably well. 
    
    \item While the radial number density profile of subhalos can be as steep as the dark matter density profile in CDM, the subhalo profile remains shallower than dark matter in SIDM even when disrupted systems are accounted for.
    
    \item The coring and enhanced stripping of subhalos prevalent in SIDM can be degenerate with the parameters that control the galaxy occupation of disrupted subhalos, e.g., the effects of surviving subhalos' cored profiles in SIDM can be mimicked by a CDM model with an enhanced orphan fraction.
    
    \item The degeneracy between coring in SIDM and orphan modelling can be broken by studying the weak lensing signal around satellite galaxies in cluster outskirts ($> 0.5r_{200}$), where disrupted subhalos are rarer, especially in CDM.
    
    \item Given state-of-the-art weak lensing covariances, large-scale systematics like cluster miscentering are important compared to baryonic effects at the galaxy center. Nonetheless, we forecast that LSST will able to constrain $\sigma/m$ at the $\sim 1\ \mathrm{cm}^2\ \mathrm{g}^{-1}$ level from satellite galaxy--galaxy weak lensing measurements.
    
 \end{enumerate}
 
N-body simulations have helped us study the evolution substructure in massive halos to a great detail. The dynamics of subhalos in clusters can potentially be complicated; the precise orbit of the subhalo determining the evolution of its internal structure through time. We find that the density profiles of subhalos, particularly those that have been disrupted, can be significantly altered from their isolated counterparts and we attempt for the first time in this work to consistently incorporate the effect of these disrupted systems on the observables both in CDM and SIDM. We also note that while the central cusp in disrupted systems survives in CDM, the central regions of SIDM subhalos do not (Appendix \ref{conv_test}); this poses a challenge to assign galaxies to SIDM subhalos by traditional methods using bound particles and in future it will be important to explore alternative methods. As massive clusters are relatively easy to observe, the rich diversity of substructure allows us to study hierarchical structure formation and understand the nature of dark matter. Developing semi-analytical treatments of galaxy evolution using N-body simulations of SIDM is therefore essential to exploit the large statistical samples that will be available to us in the near future. 
 
While in this work we have focussed primarily on bright satellites of cluster halos, in principle many of our inferences can be extended to lower mass subhalos. Recent observations help us probe the lower end of the halo mass function allowing us to study fainter systems like ultra-diffuse galaxies (UDGs;  \citealt{2015ApJ...798L..45V,2015ApJ...807L...2K,2015ApJ...809L..21M, Tanoglidis:2020wdy}. Intriguing deviations from CDM have been pointed out in \citep{2020Sci...369.1347M} where they find a potential excess of small-scale lenses in strong lensing studies, while we do not see an excess of substructure in SIDM systems in the mass range that we have explored, at lower mass scales, core collapse can potentially make substructure robust to disruption near the central region of clusters. Extending SIDM zoom-in simulations into the dwarf galaxy regime will therefore be important to study substructure in the low surface brightness regime.  

Future surveys that probe much larger volumes will give us the opportunity to probe thousands of galaxy clusters significantly reducing statistical uncertainty in lensing measurements. eROSITA \citep{Pillepich:2011zz} will provide X-ray samples that will allow us to study objects that extend down to group mass, surveys like CMB S4 \citep{Abazajian:2016yjj} and Simons Observatory \citep{Ade:2018sbj} will add to the already existing catalog of Sunyaev-Zeldovich (SZ) selected clusters from Planck \citep{Ade:2013zuv}, Atacama Cosmology Telescope \citep{Hilton:2020qsa} and the South Pole Telescope \citep{Williamson11}. Moreover, the sample of optical clusters will also largely increase with the advent of the Rubin Observatory. While accurate substructure modeling is an ongoing, open problem, our set of zoom-in cluster simulations has enabled us to develop an understanding of the differences in the distributions in surviving and disrupted subhalos when self-interactions are introduced. A further understanding of the co-evolution of baryons and dark matter in SIDM simulations to constrain the detailed galaxy--halo connection should lead to more precise predictions, which, coupled with future surveys, can be used to elucidate the nature of dark matter.


\section*{Acknowledgments}

We thank Annika Peter for her valuable comments on an early version of the draft. We also thank Andrey Kravtsov and Yuanyuan Zhang for useful discussions. SB would like to thank the High Performance Computing facility at IUCAA for allowing usage of the Pegasus cluster. SB is also grateful to IUCAA for the hospitality provided during a visit. SB received support from The Ohio State University Graduate Fellowship.

This research received support from the National Science Foundation (NSF) under grant No.\ NSF DGE-1656518 through the NSF Graduate Research Fellowship received by EON. SC acknowledges support from the Department of Science and Technology (DST), GOI, through the SERB Core Research grant. AB was supported by the Fermi Research Alliance, LLC under Contract No. DE-AC02-07CH11359 with
the U.S. Department of Energy, and the U.S. Department of Energy (DOE) Office of Science Distinguished Scientist Fellow Program.

\bibliographystyle{yahapj}
\bibliography{references}

\begin{thebibliography}{}
\providecommand\natexlab[1]{#1}
\providecommand\JournalTitle[1]{#1}

\bibitem[{Abazajian {et~al.}(2016)}]{Abazajian:2016yjj}
Abazajian, K.~N., {et~al.} 2016,
  \href{http://arxiv.org/abs/1610.02743}{{\sffamily arXiv:1610.02743
  [astro-ph.CO]}}

\bibitem[{Abbott {et~al.}(2005)}]{Abbott:2005bi}
Abbott, T., {et~al.} 2005,
  \href{http://arxiv.org/abs/astro-ph/0510346}{{\sffamily
  arXiv:astro-ph/0510346}}

\bibitem[{{Abbott} {et~al.}(2018){Abbott}, {Abdalla}, {Alarcon}, {Aleksi{\'c}},
  {Allam}, {Allen}, {Amara}, {Annis}, {Asorey}, {Avila}, {Bacon}, {Balbinot},
  {Banerji}, {Banik}, {Barkhouse}, {Baumer}, {Baxter}, {Bechtol}, {Becker},
  {Benoit-L{\'e}vy}, {Benson}, {Bernstein}, {Bertin}, {Blazek}, {Bridle},
  {Brooks}, {Brout}, {Buckley-Geer}, {Burke}, {Busha}, {Campos}, {Capozzi},
  {Carnero Rosell}, {Carrasco Kind}, {Carretero}, {Castander}, {Cawthon},
  {Chang}, {Chen}, {Childress}, {Choi}, {Conselice}, {Crittenden}, {Crocce},
  {Cunha}, {D'Andrea}, {da Costa}, {Das}, {Davis}, {Davis}, {De Vicente},
  {DePoy}, {DeRose}, {Desai}, {Diehl}, {Dietrich}, {Dodelson}, {Doel},
  {Drlica-Wagner}, {Eifler}, {Elliott}, {Elsner}, {Elvin-Poole}, {Estrada},
  {Evrard}, {Fang}, {Fernandez}, {Fert{\'e}}, {Finley}, {Flaugher}, {Fosalba},
  {Friedrich}, {Frieman}, {Garc{\'\i}a-Bellido}, {Garcia-Fernandez}, {Gatti},
  {Gaztanaga}, {Gerdes}, {Giannantonio}, {Gill}, {Glazebrook}, {Goldstein},
  {Gruen}, {Gruendl}, {Gschwend}, {Gutierrez}, {Hamilton}, {Hartley}, {Hinton},
  {Honscheid}, {Hoyle}, {Huterer}, {Jain}, {James}, {Jarvis}, {Jeltema},
  {Johnson}, {Johnson}, {Kacprzak}, {Kent}, {Kim}, {King}, {Kirk}, {Kokron},
  {Kovacs}, {Krause}, {Krawiec}, {Kremin}, {Kuehn}, {Kuhlmann}, {Kuropatkin},
  {Lacasa}, {Lahav}, {Li}, {Liddle}, {Lidman}, {Lima}, {Lin}, {MacCrann},
  {Maia}, {Makler}, {Manera}, {March}, {Marshall}, {Martini}, {McMahon},
  {Melchior}, {Menanteau}, {Miquel}, {Miranda}, {Mudd}, {Muir}, {M{\"o}ller},
  {Neilsen}, {Nichol}, {Nord}, {Nugent}, {Ogando}, {Palmese}, {Peacock},
  {Peiris}, {Peoples}, {Percival}, {Petravick}, {Plazas}, {Porredon}, {Prat},
  {Pujol}, {Rau}, {Refregier}, {Ricker}, {Roe}, {Rollins}, {Romer}, {Roodman},
  {Rosenfeld}, {Ross}, {Rozo}, {Rykoff}, {Sako}, {Salvador}, {Samuroff},
  {S{\'a}nchez}, {Sanchez}, {Santiago}, {Scarpine}, {Schindler}, {Scolnic},
  {Secco}, {Serrano}, {Sevilla-Noarbe}, {Sheldon}, {Smith}, {Smith}, {Smith},
  {Soares-Santos}, {Sobreira}, {Suchyta}, {Tarle}, {Thomas}, {Troxel},
  {Tucker}, {Tucker}, {Uddin}, {Varga}, {Vielzeuf}, {Vikram}, {Vivas},
  {Walker}, {Wang}, {Wechsler}, {Weller}, {Wester}, {Wolf}, {Yanny}, {Yuan},
  {Zenteno}, {Zhang}, {Zhang}, {Zuntz}, \& {Dark Energy Survey
  Collaboration}}]{2018PhRvD..98d3526A}
{Abbott}, T.~M.~C., {Abdalla}, F.~B., {Alarcon}, A., {et~al.} 2018,
  \href{http://dx.doi.org/10.1103/PhysRevD.98.043526}{\JournalTitle{\prd}, 98,
  043526}

\bibitem[{Abell {et~al.}(2009)}]{Abell:2009aa}
Abell, P.~A., {et~al.} 2009, \href{http://arxiv.org/abs/0912.0201}{{\sffamily
  arXiv:0912.0201 [astro-ph.IM]}}

\bibitem[{Ade {et~al.}(2014)}]{Ade:2013zuv}
Ade, P., {et~al.} 2014,
  \href{http://dx.doi.org/10.1051/0004-6361/201321591}{\JournalTitle{Astron.
  Astrophys.}, 571, A16}

\bibitem[{Ade {et~al.}(2019)}]{Ade:2018sbj}
---. 2019,
  \href{http://dx.doi.org/10.1088/1475-7516/2019/02/056}{\JournalTitle{JCAP},
  02, 056}

\bibitem[{{Aihara} {et~al.}(2018){Aihara}, {Arimoto}, {Armstrong}, {Arnouts},
  {Bahcall}, {Bickerton}, {Bosch}, {Bundy}, {Capak}, {Chan}, {Chiba}, {Coupon},
  {Egami}, {Enoki}, {Finet}, {Fujimori}, {Fujimoto}, {Furusawa}, {Furusawa},
  {Goto}, {Goulding}, {Greco}, {Greene}, {Gunn}, {Hamana}, {Harikane},
  {Hashimoto}, {Hattori}, {Hayashi}, {Hayashi}, {He{\l}miniak}, {Higuchi},
  {Hikage}, {Ho}, {Hsieh}, {Huang}, {Huang}, {Ikeda}, {Imanishi}, {Inoue},
  {Iwasawa}, {Iwata}, {Jaelani}, {Jian}, {Kamata}, {Karoji}, {Kashikawa},
  {Katayama}, {Kawanomoto}, {Kayo}, {Koda}, {Koike}, {Kojima}, {Komiyama},
  {Konno}, {Koshida}, {Koyama}, {Kusakabe}, {Leauthaud}, {Lee}, {Lin}, {Lin},
  {Lupton}, {Mand elbaum}, {Matsuoka}, {Medezinski}, {Mineo}, {Miyama},
  {Miyatake}, {Miyazaki}, {Momose}, {More}, {More}, {Moritani}, {Moriya},
  {Morokuma}, {Mukae}, {Murata}, {Murayama}, {Nagao}, {Nakata}, {Niida},
  {Niikura}, {Nishizawa}, {Obuchi}, {Oguri}, {Oishi}, {Okabe}, {Okamoto},
  {Okura}, {Ono}, {Onodera}, {Onoue}, {Osato}, {Ouchi}, {Price}, {Pyo}, {Sako},
  {Sawicki}, {Shibuya}, {Shimasaku}, {Shimono}, {Shirasaki}, {Silverman},
  {Simet}, {Speagle}, {Spergel}, {Strauss}, {Sugahara}, {Sugiyama}, {Suto},
  {Suyu}, {Suzuki}, {Tait}, {Takada}, {Takata}, {Tamura}, {Tanaka}, {Tanaka},
  {Tanaka}, {Tanaka}, {Terai}, {Terashima}, {Toba}, {Tominaga}, {Toshikawa},
  {Turner}, {Uchida}, {Uchiyama}, {Umetsu}, {Uraguchi}, {Urata}, {Usuda},
  {Utsumi}, {Wang}, {Wang}, {Wong}, {Yabe}, {Yamada}, {Yamanoi}, {Yasuda},
  {Yeh}, {Yonehara}, \& {Yuma}}]{2018PASJ...70S...4A}
{Aihara}, H., {Arimoto}, N., {Armstrong}, R., {et~al.} 2018,
  \href{http://dx.doi.org/10.1093/pasj/psx066}{\JournalTitle{\pasj}, 70, S4}

\bibitem[{{Aihara} {et~al.}(2019){Aihara}, {AlSayyad}, {Ando}, {Armstrong},
  {Bosch}, {Egami}, {Furusawa}, {Furusawa}, {Goulding}, {Harikane}, {Hikage},
  {Ho}, {Hsieh}, {Huang}, {Ikeda}, {Imanishi}, {Ito}, {Iwata}, {Jaelani},
  {Kakuma}, {Kawana}, {Kikuta}, {Kobayashi}, {Koike}, {Komiyama}, {Li},
  {Liang}, {Lin}, {Luo}, {Lupton}, {Lust}, {MacArthur}, {Matsuoka}, {Mineo},
  {Miyatake}, {Miyazaki}, {More}, {Murata}, {Namiki}, {Nishizawa}, {Oguri},
  {Okabe}, {Okamoto}, {Okura}, {Ono}, {Onodera}, {Onoue}, {Osato}, {Ouchi},
  {Shibuya}, {Strauss}, {Sugiyama}, {Suto}, {Takada}, {Takagi}, {Takata},
  {Takita}, {Tanaka}, {Terai}, {Toba}, {Uchiyama}, {Utsumi}, {Wang}, {Wang}, \&
  {Yamada}}]{2019PASJ...71..114A}
{Aihara}, H., {AlSayyad}, Y., {Ando}, M., {et~al.} 2019,
  \href{http://dx.doi.org/10.1093/pasj/psz103}{\JournalTitle{\pasj}, 71, 114}

\bibitem[{{Banerjee} {et~al.}(2020){Banerjee}, {Adhikari}, {Dalal}, {More}, \&
  {Kravtsov}}]{Banerjee190612026}
{Banerjee}, A., {Adhikari}, S., {Dalal}, N., {More}, S., \& {Kravtsov}, A.
  2020,
  \href{http://dx.doi.org/10.1088/1475-7516/2020/02/024}{\JournalTitle{\jcap},
  2020, 024}

\bibitem[{{Bechtol} {et~al.}(2019){Bechtol}, {Drlica-Wagner}, {Abazajian},
  {Abidi}, {Adhikari}, {Ali-Ha{\i}{\ensuremath{\ddot{}}}moud}, {Annis},
  {Ansarinejad}, {Armstrong}, {Asorey}, {Baccigalupi}, {Banerjee}, {Banik},
  {Bennett}, {Beutler}, {Bird}, {Birrer}, {Biswas}, {Biviano}, {Blazek},
  {Boddy}, {Bonaca}, {Borrill}, {Bose}, {Bovy}, {Frye}, {Brooks}, {Buckley},
  {Buckley-Geer}, {Bulbul}, {Burchat}, {Burgess}, {Calore}, {Caputo},
  {Castorina}, {Chang}, {Chapline}, {Charles}, {Chen}, {Clowe}, {Cohen-Tanugi},
  {Comparat}, {Croft}, {Cuoco}, {Cyr-Racine}, {D'Amico}, {Davis}, {Dawson}, {de
  la Macorra}, {Di Valentino}, {Rivero}, {Digel}, {Dodelson}, {Dor{\'e}},
  {Dvorkin}, {Eckner}, {Ellison}, {Erkal}, {Farahi}, {Fassnacht}, {Ferreira},
  {Flaugher}, {Foreman}, {Friedrich}, {Frieman}, {Garc{\'\i}a-Bellido},
  {Gawiser}, {Gerbino}, {Giannotti}, {Gill}, {Gluscevic}, {Golovich},
  {Gontcho}, {Gonz{\'a}lez-Morales}, {Grin}, {Gruen}, {Hearin}, {Hendel},
  {Hezaveh}, {Hirata}, {Hlo{\v{z}}ek}, {Horiuchi}, {Jain}, {Jee}, {Jeltema},
  {Kamionkowski}, {Kaplinghat}, {Keeley}, {Keeton}, {Khatri}, {Koposov},
  {Koushiappas}, {Kovetz}, {Lahav}, {Lam}, {Lee}, {Li}, {Liguori}, {Lin},
  {Lisanti}, {LoVerde}, {Lu}, {Mandelbaum}, {Mao}, {McDermott}, {McNanna},
  {Medford}, {Meerburg}, {Meyer}, {Mirbabayi}, {Mishra-Sharma}, {Marc}, {More},
  {Moustakas}, {Mu{\~n}oz}, {Murgia}, {Myers}, {Nadler}, {Necib}, {Newburgh},
  {Newman}, {Nord}, {Nourbakhsh}, {Nuss}, {O'Connor}, {Pace}, {Padmanabhan},
  {Palmese}, {Peiris}, {Peter}, {Piacentni}, {Plazas}, {Polin}, {Prakash},
  {Prescod-Weinstein}, {Read}, {Ritz}, {Robertson}, {Rose}, {Rosenfeld},
  {Rossi}, {Samushia}, {S{\'a}nchez}, {S{\'a}nchez-Conde}, {Schaan}, {Sehgal},
  {Senatore}, {Seo}, {Shafieloo}, {Shan}, {Shipp}, {Simon}, {Simon}, {Slatyer},
  {Slosar}, {Sridhar}, {Stebbins}, {Straniero}, {Strigari}, {Tait}, {Tollerud},
  {Troxel}, {Tyson}, {Uhlemann}, {Uren{\~n}a-L{\'o}pez}, {Verma}, {Vilalta},
  {Walter}, {Wang}, {Watson}, {Wechsler}, {Wittman}, {Xu}, {Yanny}, {Young},
  {Yu}, {Zaharijas}, {Zentner}, \& {Zuntz}}]{2019BAAS...51c.207B}
{Bechtol}, K., {Drlica-Wagner}, A., {Abazajian}, K.~N., {et~al.} 2019,
  \JournalTitle{\baas}, 51, 207

\bibitem[{{Behroozi} {et~al.}(2019){Behroozi}, {Wechsler}, {Hearin}, \&
  {Conroy}}]{2019MNRAS.488.3143B}
{Behroozi}, P., {Wechsler}, R.~H., {Hearin}, A.~P., \& {Conroy}, C. 2019,
  \href{http://dx.doi.org/10.1093/mnras/stz1182}{\JournalTitle{\mnras}, 488,
  3143}

\bibitem[{{Behroozi} {et~al.}(2013{\natexlab{a}}){Behroozi}, {Wechsler}, \&
  {Wu}}]{2013ApJ...762..109B}
{Behroozi}, P.~S., {Wechsler}, R.~H., \& {Wu}, H.-Y. 2013{\natexlab{a}},
  \href{http://dx.doi.org/10.1088/0004-637X/762/2/109}{\JournalTitle{\apj},
  762, 109}

\bibitem[{{Behroozi} {et~al.}(2013{\natexlab{b}}){Behroozi}, {Wechsler}, {Wu},
  {Busha}, {Klypin}, \& {Primack}}]{2013ApJ...763...18B}
{Behroozi}, P.~S., {Wechsler}, R.~H., {Wu}, H.-Y., {et~al.} 2013{\natexlab{b}},
  \href{http://dx.doi.org/10.1088/0004-637X/763/1/18}{\JournalTitle{\apj}, 763,
  18}

\bibitem[{{Benson} {et~al.}(2002){Benson}, {Frenk}, {Lacey}, {Baugh}, \&
  {Cole}}]{2002MNRAS.333..177B}
{Benson}, A.~J., {Frenk}, C.~S., {Lacey}, C.~G., {Baugh}, C.~M., \& {Cole}, S.
  2002,
  \href{http://dx.doi.org/10.1046/j.1365-8711.2002.05388.x}{\JournalTitle{\mnras},
  333, 177}

\bibitem[{{Binney} \& {Tremaine}(1987)}]{1987gady.book.....B}
{Binney}, J., \& {Tremaine}, S. 1987, {Galactic dynamics}

\bibitem[{{Binney} \& {Tremaine}(2008)}]{2008gady.book.....B}
---. 2008, {Galactic Dynamics: Second Edition}

\bibitem[{{Blanton} {et~al.}(2003){Blanton}, {Hogg}, {Bahcall}, {Brinkmann},
  {Britton}, {Connolly}, {Csabai}, {Fukugita}, {Loveday}, {Meiksin}, {Munn},
  {Nichol}, {Okamura}, {Quinn}, {Schneider}, {Shimasaku}, {Strauss}, {Tegmark},
  {Vogeley}, \& {Weinberg}}]{2003ApJ...592..819B}
{Blanton}, M.~R., {Hogg}, D.~W., {Bahcall}, N.~A., {et~al.} 2003,
  \href{http://dx.doi.org/10.1086/375776}{\JournalTitle{\apj}, 592, 819}

\bibitem[{{Bose} {et~al.}(2019){Bose}, {Eisenstein}, {Hernquist}, {Pillepich},
  {Nelson}, {Marinacci}, {Springel}, \& {Vogelsberger}}]{2019MNRAS.490.5693B}
{Bose}, S., {Eisenstein}, D.~J., {Hernquist}, L., {et~al.} 2019,
  \href{http://dx.doi.org/10.1093/mnras/stz2546}{\JournalTitle{\mnras}, 490,
  5693}

\bibitem[{{Brinckmann} {et~al.}(2018){Brinckmann}, {Zavala}, {Rapetti},
  {Hansen}, \& {Vogelsberger}}]{2018MNRAS.474..746B}
{Brinckmann}, T., {Zavala}, J., {Rapetti}, D., {Hansen}, S.~H., \&
  {Vogelsberger}, M. 2018,
  \href{http://dx.doi.org/10.1093/mnras/stx2782}{\JournalTitle{\mnras}, 474,
  746}

\bibitem[{{Brooks} \& {Zolotov}(2014)}]{2014ApJ...786...87B}
{Brooks}, A.~M., \& {Zolotov}, A. 2014,
  \href{http://dx.doi.org/10.1088/0004-637X/786/2/87}{\JournalTitle{\apj}, 786,
  87}

\bibitem[{Budzynski {et~al.}(2012)Budzynski, Koposov, McCarthy, McGee, \&
  Belokurov}]{Budzynski:2012yd}
Budzynski, J.~M., Koposov, S., McCarthy, I.~G., McGee, S.~L., \& Belokurov, V.
  2012,
  \href{http://dx.doi.org/10.1111/j.1365-2966.2012.20663.x}{\JournalTitle{Mon.
  Not. Roy. Astron. Soc.}, 423, 104}

\bibitem[{{Bullock} \& {Boylan-Kolchin}(2017)}]{2017ARA&A..55..343B}
{Bullock}, J.~S., \& {Boylan-Kolchin}, M. 2017,
  \href{http://dx.doi.org/10.1146/annurev-astro-091916-055313}{\JournalTitle{\araa},
  55, 343}

\bibitem[{{Burkert}(2000)}]{2000ApJ...534L.143B}
{Burkert}, A. 2000,
  \href{http://dx.doi.org/10.1086/312674}{\JournalTitle{\apjl}, 534, L143}

\bibitem[{{Campbell} {et~al.}(2018){Campbell}, {van den Bosch}, {Padmanabhan},
  {Mao}, {Zentner}, {Lange}, {Jiang}, \& {Villarreal}}]{2018MNRAS.477..359C}
{Campbell}, D., {van den Bosch}, F.~C., {Padmanabhan}, N., {et~al.} 2018,
  \href{http://dx.doi.org/10.1093/mnras/sty495}{\JournalTitle{\mnras}, 477,
  359}

\bibitem[{Chandrasekhar(1949)}]{RevModPhys.21.383}
Chandrasekhar, S. 1949,
  \href{http://dx.doi.org/10.1103/RevModPhys.21.383}{\JournalTitle{Rev. Mod.
  Phys.}, 21, 383}

\bibitem[{{Chang} {et~al.}(2013){Chang}, {Jarvis}, {Jain}, {Kahn}, {Kirkby},
  {Connolly}, {Krughoff}, {Peng}, \& {Peterson}}]{2013MNRAS.434.2121C}
{Chang}, C., {Jarvis}, M., {Jain}, B., {et~al.} 2013,
  \href{http://dx.doi.org/10.1093/mnras/stt1156}{\JournalTitle{\mnras}, 434,
  2121}

\bibitem[{{Conroy} {et~al.}(2006){Conroy}, {Wechsler}, \&
  {Kravtsov}}]{2006ApJ...647..201C}
{Conroy}, C., {Wechsler}, R.~H., \& {Kravtsov}, A.~V. 2006,
  \href{http://dx.doi.org/10.1086/503602}{\JournalTitle{\apj}, 647, 201}

\bibitem[{{Correa}(2021)}]{2021MNRAS.503..920C}
{Correa}, C.~A. 2021,
  \href{http://dx.doi.org/10.1093/mnras/stab506}{\JournalTitle{\mnras}, 503,
  920}

\bibitem[{{de Blok}(2010)}]{2010AdAst2010E...5D}
{de Blok}, W.~J.~G. 2010,
  \href{http://dx.doi.org/10.1155/2010/789293}{\JournalTitle{Advances in
  Astronomy}, 2010, 789293}

\bibitem[{{De Lucia} {et~al.}(2006){De Lucia}, {Springel}, {White}, {Croton},
  \& {Kauffmann}}]{2006MNRAS.366..499D}
{De Lucia}, G., {Springel}, V., {White}, S. D.~M., {Croton}, D., \&
  {Kauffmann}, G. 2006,
  \href{http://dx.doi.org/10.1111/j.1365-2966.2005.09879.x}{\JournalTitle{\mnras},
  366, 499}

\bibitem[{{Delfino} {et~al.}(2021){Delfino}, {Scoccola}, {Cora},
  {Vega-Martinez}, \& {Gargiulo}}]{2021arXiv210201837D}
{Delfino}, F.~M., {Scoccola}, C.~G., {Cora}, S.~A., {Vega-Martinez}, C.~A., \&
  {Gargiulo}, I.~D. 2021, \JournalTitle{arXiv e-prints}, arXiv:2102.01837

\bibitem[{Despali {et~al.}(2019)Despali, Sparre, Vegetti, Vogelsberger, Zavala,
  \& Marinacci}]{Despali:2018zpw}
Despali, G., Sparre, M., Vegetti, S., {et~al.} 2019,
  \href{http://dx.doi.org/10.1093/mnras/stz273}{\JournalTitle{Mon. Not. Roy.
  Astron. Soc.}, 484, 4563}

\bibitem[{{Dooley} {et~al.}(2016){Dooley}, {Peter}, {Vogelsberger}, {Zavala},
  \& {Frebel}}]{2016MNRAS.461..710D}
{Dooley}, G.~A., {Peter}, A. H.~G., {Vogelsberger}, M., {Zavala}, J., \&
  {Frebel}, A. 2016,
  \href{http://dx.doi.org/10.1093/mnras/stw1309}{\JournalTitle{\mnras}, 461,
  710}

\bibitem[{{Drlica-Wagner} {et~al.}(2019){Drlica-Wagner}, {Mao}, {Adhikari},
  {Armstrong}, {Banerjee}, {Banik}, {Bechtol}, {Bird}, {Boddy}, {Bonaca},
  {Bovy}, {Buckley}, {Bulbul}, {Chang}, {Chapline}, {Cohen-Tanugi}, {Cuoco},
  {Cyr-Racine}, {Dawson}, {D{\'\i}az Rivero}, {Dvorkin}, {Erkal}, {Fassnacht},
  {Garc{\'\i}a-Bellido}, {Giannotti}, {Gluscevic}, {Golovich}, {Hendel},
  {Hezaveh}, {Horiuchi}, {Jee}, {Kaplinghat}, {Keeton}, {Koposov}, {Li},
  {Mandelbaum}, {McDermott}, {McNanna}, {Medford}, {Meyer}, {Marc}, {Murgia},
  {Nadler}, {Necib}, {Nuss}, {Pace}, {Peter}, {Polin}, {Prescod- Weinstein},
  {Read}, {Rosenfeld}, {Shipp}, {Simon}, {Slatyer}, {Straniero}, {Strigari},
  {Tollerud}, {Tyson}, {Wang}, {Wechsler}, {Wittman}, {Yu}, {Zaharijas},
  {Ali-Ha{\"\i}moud}, {Annis}, {Birrer}, {Biswas}, {Blazek}, {Brooks},
  {Buckley-Geer}, {Caputo}, {Charles}, {Digel}, {Dodelson}, {Flaugher},
  {Frieman}, {Gawiser}, {Hearin}, {Hlo{\v{z}}ek}, {Jain}, {Jeltema},
  {Koushiappas}, {Lisanti}, {LoVerde}, {Mishra-Sharma}, {Newman}, {Nord},
  {Nourbakhsh}, {Ritz}, {Robertson}, {S{\'a}nchez-Conde}, {Slosar}, {Tait},
  {Verma}, {Vilalta}, {Walter}, {Yanny}, \& {Zentner}}]{Drlica-Wagner190201055}
{Drlica-Wagner}, A., {Mao}, Y.-Y., {Adhikari}, S., {et~al.} 2019,
  \JournalTitle{arXiv e-prints}, arXiv:1902.01055

\bibitem[{{Elbert} {et~al.}(2018){Elbert}, {Bullock}, {Kaplinghat},
  {Garrison-Kimmel}, {Graus}, \& {Rocha}}]{2018ApJ...853..109E}
{Elbert}, O.~D., {Bullock}, J.~S., {Kaplinghat}, M., {et~al.} 2018,
  \href{http://dx.doi.org/10.3847/1538-4357/aa9710}{\JournalTitle{\apj}, 853,
  109}

\bibitem[{{Errani} \& {Navarro}(2021)}]{2021MNRAS.tmp.1267E}
{Errani}, R., \& {Navarro}, J.~F. 2021,
  \href{http://dx.doi.org/10.1093/mnras/stab1215}{\JournalTitle{\mnras}},
  \href{http://arxiv.org/abs/2011.07077}{{\sffamily arXiv:2011.07077
  [astro-ph.GA]}}

\bibitem[{{Fitts} {et~al.}(2018){Fitts}, {Boylan-Kolchin}, {Bozek}, {Bullock},
  {Graus}, {Robles}, {Hopkins}, {El-Badry}, {Garrison-Kimmel},
  {Faucher-Gigu{\`e}re}, {Wetzel}, \& {Kere{\v{s}}}}]{Fitts181111791}
{Fitts}, A., {Boylan-Kolchin}, M., {Bozek}, B., {et~al.} 2018,
  \JournalTitle{arXiv e-prints}, arXiv:1811.11791

\bibitem[{{Fry} {et~al.}(2015){Fry}, {Governato}, {Pontzen}, {Quinn},
  {Tremmel}, {Anderson}, {Menon}, {Brooks}, \& {Wadsley}}]{Fry150100497}
{Fry}, A.~B., {Governato}, F., {Pontzen}, A., {et~al.} 2015,
  \href{http://dx.doi.org/10.1093/mnras/stv1330}{\JournalTitle{\mnras}, 452,
  1468}

\bibitem[{{Gao} {et~al.}(2004){Gao}, {De Lucia}, {White}, \&
  {Jenkins}}]{2004MNRAS.352L...1G}
{Gao}, L., {De Lucia}, G., {White}, S.~D.~M., \& {Jenkins}, A. 2004,
  \href{http://dx.doi.org/10.1111/j.1365-2966.2004.08098.x}{\JournalTitle{\mnras},
  352, L1}

\bibitem[{{Gentile} {et~al.}(2004){Gentile}, {Salucci}, {Klein}, {Vergani}, \&
  {Kalberla}}]{2004MNRAS.351..903G}
{Gentile}, G., {Salucci}, P., {Klein}, U., {Vergani}, D., \& {Kalberla}, P.
  2004,
  \href{http://dx.doi.org/10.1111/j.1365-2966.2004.07836.x}{\JournalTitle{\mnras},
  351, 903}

\bibitem[{{Gnedin}(2000)}]{2000ApJ...542..535G}
{Gnedin}, N.~Y. 2000,
  \href{http://dx.doi.org/10.1086/317042}{\JournalTitle{\apj}, 542, 535}

\bibitem[{Gnedin {et~al.}(2004)Gnedin, Kravtsov, Klypin, \&
  Nagai}]{Gnedin:2004cx}
Gnedin, O.~Y., Kravtsov, A.~V., Klypin, A.~A., \& Nagai, D. 2004,
  \href{http://dx.doi.org/10.1086/424914}{\JournalTitle{Astrophys. J.}, 616,
  16}

\bibitem[{{Gnedin} \& {Ostriker}(2001)}]{2001ApJ...561...61G}
{Gnedin}, O.~Y., \& {Ostriker}, J.~P. 2001,
  \href{http://dx.doi.org/10.1086/323211}{\JournalTitle{\apj}, 561, 61}

\bibitem[{{Green} {et~al.}(2021){Green}, {van den Bosch}, \&
  {Jiang}}]{2021arXiv210301227G}
{Green}, S.~B., {van den Bosch}, F.~C., \& {Jiang}, F. 2021,
  \JournalTitle{arXiv e-prints}, arXiv:2103.01227

\bibitem[{{Guo} {et~al.}(2011){Guo}, {White}, {Boylan-Kolchin}, {De Lucia},
  {Kauffmann}, {Lemson}, {Li}, {Springel}, \& {Weinmann}}]{2011MNRAS.413..101G}
{Guo}, Q., {White}, S., {Boylan-Kolchin}, M., {et~al.} 2011,
  \href{http://dx.doi.org/10.1111/j.1365-2966.2010.18114.x}{\JournalTitle{\mnras},
  413, 101}

\bibitem[{{Hadzhiyska} {et~al.}(2021){Hadzhiyska}, {Bose}, {Eisenstein}, \&
  {Hernquist}}]{2021MNRAS.501.1603H}
{Hadzhiyska}, B., {Bose}, S., {Eisenstein}, D., \& {Hernquist}, L. 2021,
  \href{http://dx.doi.org/10.1093/mnras/staa3776}{\JournalTitle{\mnras}, 501,
  1603}

\bibitem[{Haggar {et~al.}(2021)Haggar, Pearce, Gray, Knebe, \&
  Yepes}]{Haggar:2021sqo}
Haggar, R., Pearce, F.~R., Gray, M.~E., Knebe, A., \& Yepes, G. 2021,
  \href{http://dx.doi.org/10.1093/mnras/stab064}{\JournalTitle{Mon. Not. Roy.
  Astron. Soc.}, 502, 1191}

\bibitem[{{Han} {et~al.}(2016){Han}, {Cole}, {Frenk}, \&
  {Jing}}]{2016MNRAS.457.1208H}
{Han}, J., {Cole}, S., {Frenk}, C.~S., \& {Jing}, Y. 2016,
  \href{http://dx.doi.org/10.1093/mnras/stv2900}{\JournalTitle{\mnras}, 457,
  1208}

\bibitem[{{Harvey} {et~al.}(2019){Harvey}, {Robertson}, {Massey}, \&
  {McCarthy}}]{2019MNRAS.488.1572H}
{Harvey}, D., {Robertson}, A., {Massey}, R., \& {McCarthy}, I.~G. 2019,
  \href{http://dx.doi.org/10.1093/mnras/stz1816}{\JournalTitle{\mnras}, 488,
  1572}

\bibitem[{Hilton {et~al.}(2021)}]{Hilton:2020qsa}
Hilton, M., {et~al.} 2021,
  \href{http://dx.doi.org/10.3847/1538-4365/abd023}{\JournalTitle{Astrophys. J.
  Suppl.}, 253, 3}

\bibitem[{{Huo} {et~al.}(2018){Huo}, {Kaplinghat}, {Pan}, \&
  {Yu}}]{2018PhLB..783...76H}
{Huo}, R., {Kaplinghat}, M., {Pan}, Z., \& {Yu}, H.-B. 2018,
  \href{http://dx.doi.org/10.1016/j.physletb.2018.06.024}{\JournalTitle{Physics
  Letters B}, 783, 76}

\bibitem[{{Ibe} \& {Yu}(2010)}]{Ibe09125425}
{Ibe}, M., \& {Yu}, H.-B. 2010,
  \href{http://dx.doi.org/10.1016/j.physletb.2010.07.026}{\JournalTitle{Physics
  Letters B}, 692, 70}

\bibitem[{{Ivezi{\'c}} {et~al.}(2019){Ivezi{\'c}}, {Kahn}, {Tyson}, {Abel},
  {Acosta}, {Allsman}, {Alonso}, {AlSayyad}, {Anderson}, {Andrew}, {Angel},
  {Angeli}, {Ansari}, {Antilogus}, {Araujo}, {Armstrong}, {Arndt}, {Astier},
  {Aubourg}, {Auza}, {Axelrod}, {Bard}, {Barr}, {Barrau}, {Bartlett}, {Bauer},
  {Bauman}, {Baumont}, {Bechtol}, {Bechtol}, {Becker}, {Becla}, {Beldica},
  {Bellavia}, {Bianco}, {Biswas}, {Blanc}, {Blazek}, {Blandford}, {Bloom},
  {Bogart}, {Bond}, {Booth}, {Borgland}, {Borne}, {Bosch}, {Boutigny},
  {Brackett}, {Bradshaw}, {Brandt}, {Brown}, {Bullock}, {Burchat}, {Burke},
  {Cagnoli}, {Calabrese}, {Callahan}, {Callen}, {Carlin}, {Carlson},
  {Chandrasekharan}, {Charles-Emerson}, {Chesley}, {Cheu}, {Chiang}, {Chiang},
  {Chirino}, {Chow}, {Ciardi}, {Claver}, {Cohen-Tanugi}, {Cockrum}, {Coles},
  {Connolly}, {Cook}, {Cooray}, {Covey}, {Cribbs}, {Cui}, {Cutri}, {Daly},
  {Daniel}, {Daruich}, {Daubard}, {Daues}, {Dawson}, {Delgado}, {Dellapenna},
  {de Peyster}, {de Val-Borro}, {Digel}, {Doherty}, {Dubois},
  {Dubois-Felsmann}, {Durech}, {Economou}, {Eifler}, {Eracleous}, {Emmons},
  {Fausti Neto}, {Ferguson}, {Figueroa}, {Fisher-Levine}, {Focke}, {Foss},
  {Frank}, {Freemon}, {Gangler}, {Gawiser}, {Geary}, {Gee}, {Geha}, {Gessner},
  {Gibson}, {Gilmore}, {Glanzman}, {Glick}, {Goldina}, {Goldstein}, {Goodenow},
  {Graham}, {Gressler}, {Gris}, {Guy}, {Guyonnet}, {Haller}, {Harris},
  {Hascall}, {Haupt}, {Hernandez}, {Herrmann}, {Hileman}, {Hoblitt}, {Hodgson},
  {Hogan}, {Howard}, {Huang}, {Huffer}, {Ingraham}, {Innes}, {Jacoby}, {Jain},
  {Jammes}, {Jee}, {Jenness}, {Jernigan}, {Jevremovi{\'c}}, {Johns}, {Johnson},
  {Johnson}, {Jones}, {Juramy-Gilles}, {Juri{\'c}}, {Kalirai}, {Kallivayalil},
  {Kalmbach}, {Kantor}, {Karst}, {Kasliwal}, {Kelly}, {Kessler}, {Kinnison},
  {Kirkby}, {Knox}, {Kotov}, {Krabbendam}, {Krughoff}, {Kub{\'a}nek},
  {Kuczewski}, {Kulkarni}, {Ku}, {Kurita}, {Lage}, {Lambert}, {Lange},
  {Langton}, {Le Guillou}, {Levine}, {Liang}, {Lim}, {Lintott}, {Long},
  {Lopez}, {Lotz}, {Lupton}, {Lust}, {MacArthur}, {Mahabal}, {Mandelbaum},
  {Markiewicz}, {Marsh}, {Marshall}, {Marshall}, {May}, {McKercher}, {McQueen},
  {Meyers}, {Migliore}, {Miller}, {Mills}, {Miraval}, {Moeyens}, {Moolekamp},
  {Monet}, {Moniez}, {Monkewitz}, {Montgomery}, {Morrison}, {Mueller},
  {Muller}, {Mu{\~n}oz Arancibia}, {Neill}, {Newbry}, {Nief}, {Nomerotski},
  {Nordby}, {O'Connor}, {Oliver}, {Olivier}, {Olsen}, {O'Mullane}, {Ortiz},
  {Osier}, {Owen}, {Pain}, {Palecek}, {Parejko}, {Parsons}, {Pease},
  {Peterson}, {Peterson}, {Petravick}, {Libby Petrick}, {Petry},
  {Pierfederici}, {Pietrowicz}, {Pike}, {Pinto}, {Plante}, {Plate}, {Plutchak},
  {Price}, {Prouza}, {Radeka}, {Rajagopal}, {Rasmussen}, {Regnault}, {Reil},
  {Reiss}, {Reuter}, {Ridgway}, {Riot}, {Ritz}, {Robinson}, {Roby}, {Roodman},
  {Rosing}, {Roucelle}, {Rumore}, {Russo}, {Saha}, {Sassolas}, {Schalk},
  {Schellart}, {Schindler}, {Schmidt}, {Schneider}, {Schneider}, {Schoening},
  {Schumacher}, {Schwamb}, {Sebag}, {Selvy}, {Sembroski}, {Seppala}, {Serio},
  {Serrano}, {Shaw}, {Shipsey}, {Sick}, {Silvestri}, {Slater}, {Smith},
  {Smith}, {Sobhani}, {Soldahl}, {Storrie-Lombardi}, {Stover}, {Strauss},
  {Street}, {Stubbs}, {Sullivan}, {Sweeney}, {Swinbank}, {Szalay}, {Takacs},
  {Tether}, {Thaler}, {Thayer}, {Thomas}, {Thornton}, {Thukral}, {Tice},
  {Trilling}, {Turri}, {Van Berg}, {Vanden Berk}, {Vetter}, {Virieux},
  {Vucina}, {Wahl}, {Walkowicz}, {Walsh}, {Walter}, {Wang}, {Wang}, {Warner},
  {Wiecha}, {Willman}, {Winters}, {Wittman}, {Wolff}, {Wood-Vasey}, {Wu},
  {Xin}, {Yoachim}, \& {Zhan}}]{2019ApJ...873..111I}
{Ivezi{\'c}}, {\v{Z}}., {Kahn}, S.~M., {Tyson}, J.~A., {et~al.} 2019,
  \href{http://dx.doi.org/10.3847/1538-4357/ab042c}{\JournalTitle{\apj}, 873,
  111}

\bibitem[{{Jiang} {et~al.}(2021){Jiang}, {Dekel}, {Freundlich}, {van den
  Bosch}, {Green}, {Hopkins}, {Benson}, \& {Du}}]{2021MNRAS.502..621J}
{Jiang}, F., {Dekel}, A., {Freundlich}, J., {et~al.} 2021,
  \href{http://dx.doi.org/10.1093/mnras/staa4034}{\JournalTitle{\mnras}, 502,
  621}

\bibitem[{{Johnston} {et~al.}(2007){Johnston}, {Sheldon}, {Wechsler}, {Rozo},
  {Koester}, {Frieman}, {McKay}, {Evrard}, {Becker}, \&
  {Annis}}]{2007arXiv0709.1159J}
{Johnston}, D.~E., {Sheldon}, E.~S., {Wechsler}, R.~H., {et~al.} 2007,
  \JournalTitle{arXiv e-prints}, arXiv:0709.1159

\bibitem[{Kahlhoefer {et~al.}(2014)Kahlhoefer, Schmidt-Hoberg, Frandsen, \&
  Sarkar}]{Kahlhoefer:2013dca}
Kahlhoefer, F., Schmidt-Hoberg, K., Frandsen, M.~T., \& Sarkar, S. 2014,
  \href{http://dx.doi.org/10.1093/mnras/stt2097}{\JournalTitle{Mon. Not. Roy.
  Astron. Soc.}, 437, 2865}

\bibitem[{Kaplinghat {et~al.}(2014)Kaplinghat, Keeley, Linden, \&
  Yu}]{Kaplinghat:2013xca}
Kaplinghat, M., Keeley, R.~E., Linden, T., \& Yu, H.-B. 2014,
  \href{http://dx.doi.org/10.1103/PhysRevLett.113.021302}{\JournalTitle{Phys.
  Rev. Lett.}, 113, 021302}

\bibitem[{{Kaplinghat} {et~al.}(2016){Kaplinghat}, {Tulin}, \&
  {Yu}}]{Kaplinghat158003339}
{Kaplinghat}, M., {Tulin}, S., \& {Yu}, H.-B. 2016,
  \href{http://dx.doi.org/10.1103/PhysRevLett.116.041302}{\JournalTitle{\prl},
  116, 041302}

\bibitem[{Kaplinghat {et~al.}(2016)Kaplinghat, Tulin, \&
  Yu}]{Kaplinghat:2015aga}
Kaplinghat, M., Tulin, S., \& Yu, H.-B. 2016,
  \href{http://dx.doi.org/10.1103/PhysRevLett.116.041302}{\JournalTitle{Phys.
  Rev. Lett.}, 116, 041302}

\bibitem[{{Kaplinghat} {et~al.}(2019){Kaplinghat}, {Valli}, \&
  {Yu}}]{Kaplinghat190404939}
{Kaplinghat}, M., {Valli}, M., \& {Yu}, H.-B. 2019, \JournalTitle{arXiv
  e-prints}, arXiv:1904.04939

\bibitem[{Kim {et~al.}(2018)Kim, Peter, \& Hargis}]{Kim1812121}
Kim, S.~Y., Peter, A. H.~G., \& Hargis, J.~R. 2018,
  \href{http://dx.doi.org/10.1103/PhysRevLett.121.211302}{\JournalTitle{\prl},
  121, 211302}

\bibitem[{{Kim} {et~al.}(2017){Kim}, {Peter}, \&
  {Wittman}}]{2017MNRAS.469.1414K}
{Kim}, S.~Y., {Peter}, A. H.~G., \& {Wittman}, D. 2017,
  \href{http://dx.doi.org/10.1093/mnras/stx896}{\JournalTitle{\mnras}, 469,
  1414}

\bibitem[{{Klypin} {et~al.}(1999){Klypin}, {Kravtsov}, {Valenzuela}, \&
  {Prada}}]{Klypin9901240}
{Klypin}, A., {Kravtsov}, A.~V., {Valenzuela}, O., \& {Prada}, F. 1999,
  \href{http://dx.doi.org/10.1086/307643}{\JournalTitle{\apj}, 522, 82}

\bibitem[{{Koda} {et~al.}(2015){Koda}, {Yagi}, {Yamanoi}, \&
  {Komiyama}}]{2015ApJ...807L...2K}
{Koda}, J., {Yagi}, M., {Yamanoi}, H., \& {Komiyama}, Y. 2015,
  \href{http://dx.doi.org/10.1088/2041-8205/807/1/L2}{\JournalTitle{\apjl},
  807, L2}

\bibitem[{{Kravtsov} \& {Borgani}(2012)}]{2012ARA&A..50..353K}
{Kravtsov}, A.~V., \& {Borgani}, S. 2012,
  \href{http://dx.doi.org/10.1146/annurev-astro-081811-125502}{\JournalTitle{\araa},
  50, 353}

\bibitem[{{Kuhlen} {et~al.}(2012){Kuhlen}, {Vogelsberger}, \&
  {Angulo}}]{2012PDU.....1...50K}
{Kuhlen}, M., {Vogelsberger}, M., \& {Angulo}, R. 2012,
  \href{http://dx.doi.org/10.1016/j.dark.2012.10.002}{\JournalTitle{Physics of
  the Dark Universe}, 1, 50}

\bibitem[{{Kumar \& More}(in prep)}]{KumarMoreinprep}
{Kumar \& More}. in prep

\bibitem[{{Kummer} {et~al.}(2018){Kummer}, {Kahlhoefer}, \&
  {Schmidt-Hoberg}}]{Kummer170604794}
{Kummer}, J., {Kahlhoefer}, F., \& {Schmidt-Hoberg}, K. 2018,
  \href{http://dx.doi.org/10.1093/mnras/stx2715}{\JournalTitle{\mnras}, 474,
  388}

\bibitem[{{Li} {et~al.}(2014){Li}, {Shan}, {Mo}, {Kneib}, {Yang}, {Luo}, {van
  den Bosch}, {Erben}, {Moraes}, \& {Makler}}]{2014MNRAS.438.2864L}
{Li}, R., {Shan}, H., {Mo}, H., {et~al.} 2014,
  \href{http://dx.doi.org/10.1093/mnras/stt2395}{\JournalTitle{\mnras}, 438,
  2864}

\bibitem[{{Mandelbaum} {et~al.}(2008){Mandelbaum}, {Seljak}, \&
  {Hirata}}]{2008JCAP...08..006M}
{Mandelbaum}, R., {Seljak}, U., \& {Hirata}, C.~M. 2008,
  \href{http://dx.doi.org/10.1088/1475-7516/2008/08/006}{\JournalTitle{\jcap},
  2008, 006}

\bibitem[{{Mandelbaum} {et~al.}(2005){Mandelbaum}, {Hirata}, {Seljak}, {Guzik},
  {Padmanabhan}, {Blake}, {Blanton}, {Lupton}, \&
  {Brinkmann}}]{2005MNRAS.361.1287M}
{Mandelbaum}, R., {Hirata}, C.~M., {Seljak}, U., {et~al.} 2005,
  \href{http://dx.doi.org/10.1111/j.1365-2966.2005.09282.x}{\JournalTitle{\mnras},
  361, 1287}

\bibitem[{{Mandelbaum} {et~al.}(2018{\natexlab{a}}){Mandelbaum}, {Miyatake},
  {Hamana}, {Oguri}, {Simet}, {Armstrong}, {Bosch}, {Murata}, {Lanusse},
  {Leauthaud}, {Coupon}, {More}, {Takada}, {Miyazaki}, {Speagle}, {Shirasaki},
  {Sif{\'o}n}, {Huang}, {Nishizawa}, {Medezinski}, {Okura}, {Okabe}, {Czakon},
  {Takahashi}, {Coulton}, {Hikage}, {Komiyama}, {Lupton}, {Strauss}, {Tanaka},
  \& {Utsumi}}]{2018PASJ...70S..25M}
{Mandelbaum}, R., {Miyatake}, H., {Hamana}, T., {et~al.} 2018{\natexlab{a}},
  \href{http://dx.doi.org/10.1093/pasj/psx130}{\JournalTitle{\pasj}, 70, S25}

\bibitem[{{Mandelbaum} {et~al.}(2018{\natexlab{b}}){Mandelbaum}, {Lanusse},
  {Leauthaud}, {Armstrong}, {Simet}, {Miyatake}, {Meyers}, {Bosch}, {Murata},
  {Miyazaki}, \& {Tanaka}}]{2018MNRAS.481.3170M}
{Mandelbaum}, R., {Lanusse}, F., {Leauthaud}, A., {et~al.} 2018{\natexlab{b}},
  \href{http://dx.doi.org/10.1093/mnras/sty2420}{\JournalTitle{\mnras}, 481,
  3170}

\bibitem[{{Markevitch} {et~al.}(2004){Markevitch}, {Gonzalez}, {Clowe},
  {Vikhlinin}, {Forman}, {Jones}, {Murray}, \& {Tucker}}]{2004ApJ...606..819M}
{Markevitch}, M., {Gonzalez}, A.~H., {Clowe}, D., {et~al.} 2004,
  \href{http://dx.doi.org/10.1086/383178}{\JournalTitle{\apj}, 606, 819}

\bibitem[{Markevitch {et~al.}(2004)Markevitch, Gonzalez, Clowe, Vikhlinin,
  David, Forman, Jones, Murray, \& Tucker}]{Markevitch:2003at}
Markevitch, M., Gonzalez, A.~H., Clowe, D., {et~al.} 2004,
  \href{http://dx.doi.org/10.1086/383178}{\JournalTitle{Astrophys. J.}, 606,
  819}

\bibitem[{{Melchior} {et~al.}(2017){Melchior}, {Gruen}, {McClintock}, {Varga},
  {Sheldon}, {Rozo}, {Amara}, {Becker}, {Benson}, {Bermeo}, {Bridle},
  {Clampitt}, {Dietrich}, {Hartley}, {Hollowood}, {Jain}, {Jarvis}, {Jeltema},
  {Kacprzak}, {MacCrann}, {Rykoff}, {Saro}, {Suchyta}, {Troxel}, {Zuntz},
  {Bonnett}, {Plazas}, {Abbott}, {Abdalla}, {Annis}, {Benoit-L{\'e}vy},
  {Bernstein}, {Bertin}, {Brooks}, {Buckley-Geer}, {Carnero Rosell}, {Carrasco
  Kind}, {Carretero}, {Cunha}, {D'Andrea}, {da Costa}, {Desai}, {Eifler},
  {Flaugher}, {Fosalba}, {Garc{\'\i}a-Bellido}, {Gaztanaga}, {Gerdes},
  {Gruendl}, {Gschwend}, {Gutierrez}, {Honscheid}, {James}, {Kirk}, {Krause},
  {Kuehn}, {Kuropatkin}, {Lahav}, {Lima}, {Maia}, {March}, {Martini},
  {Menanteau}, {Miller}, {Miquel}, {Mohr}, {Nichol}, {Ogando}, {Romer},
  {Sanchez}, {Scarpine}, {Sevilla-Noarbe}, {Smith}, {Soares-Santos},
  {Sobreira}, {Swanson}, {Tarle}, {Thomas}, {Walker}, {Weller}, \&
  {Zhang}}]{2017MNRAS.469.4899M}
{Melchior}, P., {Gruen}, D., {McClintock}, T., {et~al.} 2017,
  \href{http://dx.doi.org/10.1093/mnras/stx1053}{\JournalTitle{\mnras}, 469,
  4899}

\bibitem[{{Meneghetti} {et~al.}(2020){Meneghetti}, {Davoli}, {Bergamini},
  {Rosati}, {Natarajan}, {Giocoli}, {Caminha}, {Metcalf}, {Rasia}, {Borgani},
  {Calura}, {Grillo}, {Mercurio}, \& {Vanzella}}]{2020Sci...369.1347M}
{Meneghetti}, M., {Davoli}, G., {Bergamini}, P., {et~al.} 2020,
  \href{http://dx.doi.org/10.1126/science.aax5164}{\JournalTitle{Science}, 369,
  1347}

\bibitem[{{Merritt}(1983)}]{1983ApJ...264...24M}
{Merritt}, D. 1983,
  \href{http://dx.doi.org/10.1086/160571}{\JournalTitle{\apj}, 264, 24}

\bibitem[{{Mihos} {et~al.}(2015){Mihos}, {Durrell}, {Ferrarese}, {Feldmeier},
  {C{\^o}t{\'e}}, {Peng}, {Harding}, {Liu}, {Gwyn}, \&
  {Cuillandre}}]{2015ApJ...809L..21M}
{Mihos}, J.~C., {Durrell}, P.~R., {Ferrarese}, L., {et~al.} 2015,
  \href{http://dx.doi.org/10.1088/2041-8205/809/2/L21}{\JournalTitle{\apjl},
  809, L21}

\bibitem[{{Miralda-Escud{\'e}}(2002)}]{2002ApJ...564...60M}
{Miralda-Escud{\'e}}, J. 2002,
  \href{http://dx.doi.org/10.1086/324138}{\JournalTitle{\apj}, 564, 60}

\bibitem[{{Moore} {et~al.}(2000){Moore}, {Gelato}, {Jenkins}, {Pearce}, \&
  {Quilis}}]{2000ApJ...535L..21M}
{Moore}, B., {Gelato}, S., {Jenkins}, A., {Pearce}, F.~R., \& {Quilis}, V.
  2000, \href{http://dx.doi.org/10.1086/312692}{\JournalTitle{\apjl}, 535, L21}

\bibitem[{{Moore} {et~al.}(1999){Moore}, {Ghigna}, {Governato}, {Lake},
  {Quinn}, {Stadel}, \& {Tozzi}}]{Moore9907411}
{Moore}, B., {Ghigna}, S., {Governato}, F., {et~al.} 1999,
  \href{http://dx.doi.org/10.1086/312287}{\JournalTitle{\apjl}, 524, L19}

\bibitem[{{More} {et~al.}(2016){More}, {Miyatake}, {Takada}, {Diemer},
  {Kravtsov}, {Dalal}, {More}, {Murata}, {Mandelbaum}, {Rozo}, {Rykoff},
  {Oguri}, \& {Spergel}}]{More16}
{More}, S., {Miyatake}, H., {Takada}, M., {et~al.} 2016,
  \href{http://dx.doi.org/10.3847/0004-637X/825/1/39}{\JournalTitle{\apj}, 825,
  39}

\bibitem[{{Nadler} {et~al.}(2020{\natexlab{a}}){Nadler}, {Banerjee},
  {Adhikari}, {Mao}, \& {Wechsler}}]{2020ApJ...896..112N}
{Nadler}, E.~O., {Banerjee}, A., {Adhikari}, S., {Mao}, Y.-Y., \& {Wechsler},
  R.~H. 2020{\natexlab{a}},
  \href{http://dx.doi.org/10.3847/1538-4357/ab94b0}{\JournalTitle{\apj}, 896,
  112}

\bibitem[{{Nadler} {et~al.}(in prep){Nadler}, {Banerjee}, {Adhikari}, {Mao}, \&
  {Wechsler}}]{Nadleretalinprep}
---. in prep

\bibitem[{Nadler {et~al.}(2019)Nadler, Mao, Green, \&
  Wechsler}]{Nadler180905542}
Nadler, E.~O., Mao, Y.-Y., Green, G.~M., \& Wechsler, R.~H. 2019,
  \href{http://dx.doi.org/10.3847/1538-4357/ab040e}{\JournalTitle{\apj}, 873,
  34}

\bibitem[{{Nadler} {et~al.}(2018){Nadler}, {Mao}, {Wechsler},
  {Garrison-Kimmel}, \& {Wetzel}}]{Nadler171204467}
{Nadler}, E.~O., {Mao}, Y.-Y., {Wechsler}, R.~H., {Garrison-Kimmel}, S., \&
  {Wetzel}, A. 2018,
  \href{http://dx.doi.org/10.3847/1538-4357/aac266}{\JournalTitle{\apj}, 859,
  129}

\bibitem[{{Nadler} {et~al.}(2020{\natexlab{b}}){Nadler}, {Wechsler}, {Bechtol},
  {Mao}, {Green}, {Drlica-Wagner}, {McNanna}, {Mau}, {Pace}, {Simon},
  {Kravtsov}, {Dodelson}, {Li}, {Riley}, {Wang}, {Abbott}, {Aguena}, {Allam},
  {Annis}, {Avila}, {Bernstein}, {Bertin}, {Brooks}, {Burke}, {Rosell}, {Kind},
  {Carretero}, {Costanzi}, {da Costa}, {De Vicente}, {Desai}, {Evrard},
  {Flaugher}, {Fosalba}, {Frieman}, {Garc{\'\i}a-Bellido}, {Gaztanaga},
  {Gerdes}, {Gruen}, {Gschwend}, {Gutierrez}, {Hartley}, {Hinton}, {Honscheid},
  {Krause}, {Kuehn}, {Kuropatkin}, {Lahav}, {Maia}, {Marshall}, {Menanteau},
  {Miquel}, {Palmese}, {Paz-Chinch{\'o}n}, {Plazas}, {Romer}, {Sanchez},
  {Santiago}, {Scarpine}, {Serrano}, {Smith}, {Soares-Santos}, {Suchyta},
  {Tarle}, {Thomas}, {Varga}, {Walker}, \& {DES
  Collaboration}}]{2020ApJ...893...48N}
{Nadler}, E.~O., {Wechsler}, R.~H., {Bechtol}, K., {et~al.} 2020{\natexlab{b}},
  \href{http://dx.doi.org/10.3847/1538-4357/ab846a}{\JournalTitle{\apj}, 893,
  48}

\bibitem[{{Natarajan} {et~al.}(2007){Natarajan}, {De Lucia}, \&
  {Springel}}]{2007MNRAS.376..180N}
{Natarajan}, P., {De Lucia}, G., \& {Springel}, V. 2007,
  \href{http://dx.doi.org/10.1111/j.1365-2966.2007.11399.x}{\JournalTitle{\mnras},
  376, 180}

\bibitem[{{Natarajan} \& {Springel}(2004)}]{2004ApJ...617L..13N}
{Natarajan}, P., \& {Springel}, V. 2004,
  \href{http://dx.doi.org/10.1086/427079}{\JournalTitle{\apjl}, 617, L13}

\bibitem[{{Natarajan} {et~al.}(2017){Natarajan}, {Chadayammuri}, {Jauzac},
  {Richard}, {Kneib}, {Ebeling}, {Jiang}, {van den Bosch}, {Limousin}, {Jullo},
  {Atek}, {Pillepich}, {Popa}, {Marinacci}, {Hernquist}, {Meneghetti}, \&
  {Vogelsberger}}]{2017MNRAS.468.1962N}
{Natarajan}, P., {Chadayammuri}, U., {Jauzac}, M., {et~al.} 2017,
  \href{http://dx.doi.org/10.1093/mnras/stw3385}{\JournalTitle{\mnras}, 468,
  1962}

\bibitem[{{Newman} {et~al.}(2013){Newman}, {Treu}, {Ellis}, {Sand}, {Nipoti},
  {Richard}, \& {Jullo}}]{2013ApJ...765...24N}
{Newman}, A.~B., {Treu}, T., {Ellis}, R.~S., {et~al.} 2013,
  \href{http://dx.doi.org/10.1088/0004-637X/765/1/24}{\JournalTitle{\apj}, 765,
  24}

\bibitem[{{Niemiec} {et~al.}(2019){Niemiec}, {Jullo}, {Giocoli}, {Limousin}, \&
  {Jauzac}}]{2019MNRAS.487..653N}
{Niemiec}, A., {Jullo}, E., {Giocoli}, C., {Limousin}, M., \& {Jauzac}, M.
  2019, \href{http://dx.doi.org/10.1093/mnras/stz1318}{\JournalTitle{\mnras},
  487, 653}

\bibitem[{{Nishikawa} {et~al.}(2019){Nishikawa}, {Boddy}, \&
  {Kaplinghat}}]{Nishikawa190100499}
{Nishikawa}, H., {Boddy}, K.~K., \& {Kaplinghat}, M. 2019, \JournalTitle{arXiv
  e-prints}, arXiv:1901.00499

\bibitem[{{Pe{\~n}arrubia} {et~al.}(2010){Pe{\~n}arrubia}, {Benson}, {Walker},
  {Gilmore}, {McConnachie}, \& {Mayer}}]{2010MNRAS.406.1290P}
{Pe{\~n}arrubia}, J., {Benson}, A.~J., {Walker}, M.~G., {et~al.} 2010,
  \href{http://dx.doi.org/10.1111/j.1365-2966.2010.16762.x}{\JournalTitle{\mnras},
  406, 1290}

\bibitem[{{Peirani} {et~al.}(2017){Peirani}, {Dubois}, {Volonteri},
  {Devriendt}, {Bundy}, {Silk}, {Pichon}, {Kaviraj}, {Gavazzi}, \&
  {Habouzit}}]{2017MNRAS.472.2153P}
{Peirani}, S., {Dubois}, Y., {Volonteri}, M., {et~al.} 2017,
  \href{http://dx.doi.org/10.1093/mnras/stx2099}{\JournalTitle{\mnras}, 472,
  2153}

\bibitem[{{Peter} {et~al.}(2013){Peter}, {Rocha}, {Bullock}, \&
  {Kaplinghat}}]{2013MNRAS.430..105P}
{Peter}, A. H.~G., {Rocha}, M., {Bullock}, J.~S., \& {Kaplinghat}, M. 2013,
  \href{http://dx.doi.org/10.1093/mnras/sts535}{\JournalTitle{\mnras}, 430,
  105}

\bibitem[{Pillepich {et~al.}(2012)Pillepich, Porciani, \&
  Reiprich}]{Pillepich:2011zz}
Pillepich, A., Porciani, C., \& Reiprich, T.~H. 2012,
  \href{http://dx.doi.org/10.1111/j.1365-2966.2012.20443.x}{\JournalTitle{Mon.
  Not. Roy. Astron. Soc.}, 422, 44}

\bibitem[{{Pontzen} \& {Governato}(2012)}]{2012MNRAS.421.3464P}
{Pontzen}, A., \& {Governato}, F. 2012,
  \href{http://dx.doi.org/10.1111/j.1365-2966.2012.20571.x}{\JournalTitle{\mnras},
  421, 3464}

\bibitem[{{Pujol} {et~al.}(2017){Pujol}, {Skibba}, {Gazta{\~n}aga}, {Benson},
  {Blaizot}, {Bower}, {Carretero}, {Castander}, {Cattaneo}, {Cora}, {Croton},
  {Cui}, {Cunnama}, {De Lucia}, {Devriendt}, {Elahi}, {Font}, {Fontanot},
  {Garcia-Bellido}, {Gargiulo}, {Gonzalez-Perez}, {Helly}, {Henriques},
  {Hirschmann}, {Knebe}, {Lee}, {Mamon}, {Monaco}, {Onions}, {Padilla},
  {Pearce}, {Power}, {Somerville}, {Srisawat}, {Thomas}, {Tollet},
  {Vega-Mart{\'\i}nez}, \& {Yi}}]{2017MNRAS.469..749P}
{Pujol}, A., {Skibba}, R.~A., {Gazta{\~n}aga}, E., {et~al.} 2017,
  \href{http://dx.doi.org/10.1093/mnras/stx913}{\JournalTitle{\mnras}, 469,
  749}

\bibitem[{{Robertson} {et~al.}(2017){Robertson}, {Massey}, \&
  {Eke}}]{2017MNRAS.467.4719R}
{Robertson}, A., {Massey}, R., \& {Eke}, V. 2017,
  \href{http://dx.doi.org/10.1093/mnras/stx463}{\JournalTitle{\mnras}, 467,
  4719}

\bibitem[{Robertson {et~al.}(2017)Robertson, Massey, \&
  Eke}]{Robertson:2016xjh}
Robertson, A., Massey, R., \& Eke, V. 2017,
  \href{http://dx.doi.org/10.1093/mnras/stw2670}{\JournalTitle{Mon. Not. Roy.
  Astron. Soc.}, 465, 569}

\bibitem[{Robertson {et~al.}(2021)Robertson, Massey, Eke, Schaye, \&
  Theuns}]{Robertson:2020pxj}
Robertson, A., Massey, R., Eke, V., Schaye, J., \& Theuns, T. 2021,
  \href{http://dx.doi.org/10.1093/mnras/staa3954}{\JournalTitle{Mon. Not. Roy.
  Astron. Soc.}, 501, 4610}

\bibitem[{{Robles} {et~al.}(2019){Robles}, {Kelley}, {Bullock}, \&
  {Kaplinghat}}]{2019MNRAS.490.2117R}
{Robles}, V.~H., {Kelley}, T., {Bullock}, J.~S., \& {Kaplinghat}, M. 2019,
  \href{http://dx.doi.org/10.1093/mnras/stz2345}{\JournalTitle{\mnras}, 490,
  2117}

\bibitem[{{Robles} {et~al.}(2017){Robles}, {Bullock}, {Elbert}, {Fitts},
  {Gonz{\'a}lez-Samaniego}, {Boylan-Kolchin}, {Hopkins}, {Faucher-Gigu{\`e}re},
  {Kere{\v{s}}}, \& {Hayward}}]{Robles170607514}
{Robles}, V.~H., {Bullock}, J.~S., {Elbert}, O.~D., {et~al.} 2017,
  \href{http://dx.doi.org/10.1093/mnras/stx2253}{\JournalTitle{\mnras}, 472,
  2945}

\bibitem[{{Rocha} {et~al.}(2013){Rocha}, {Peter}, {Bullock}, {Kaplinghat},
  {Garrison-Kimmel}, {Onorbe}, \& {Moustakas}}]{2013MNRAS.430...81R}
{Rocha}, M., {Peter}, A. H.~G., {Bullock}, J.~S., {et~al.} 2013,
  \href{http://dx.doi.org/10.1093/mnras/sts514}{\JournalTitle{\mnras}, 430, 81}

\bibitem[{{Rykoff} {et~al.}(2016){Rykoff}, {Rozo}, {Hollowood},
  {Bermeo-Hernandez}, {Jeltema}, {Mayers}, {Romer}, {Rooney}, {Saro}, {Vergara
  Cervantes}, {Wechsler}, {Wilcox}, {Abbott}, {Abdalla}, {Allam}, {Annis},
  {Benoit-L{\'e}vy}, {Bernstein}, {Bertin}, {Brooks}, {Burke}, {Capozzi},
  {Carnero Rosell}, {Carrasco Kind}, {Castander}, {Childress}, {Collins},
  {Cunha}, {D'Andrea}, {da Costa}, {Davis}, {Desai}, {Diehl}, {Dietrich},
  {Doel}, {Evrard}, {Finley}, {Flaugher}, {Fosalba}, {Frieman}, {Glazebrook},
  {Goldstein}, {Gruen}, {Gruendl}, {Gutierrez}, {Hilton}, {Honscheid}, {Hoyle},
  {James}, {Kay}, {Kuehn}, {Kuropatkin}, {Lahav}, {Lewis}, {Lidman}, {Lima},
  {Maia}, {Mann}, {Marshall}, {Martini}, {Melchior}, {Miller}, {Miquel},
  {Mohr}, {Nichol}, {Nord}, {Ogando}, {Plazas}, {Reil}, {Sahl{\'e}n},
  {Sanchez}, {Santiago}, {Scarpine}, {Schubnell}, {Sevilla-Noarbe}, {Smith},
  {Soares-Santos}, {Sobreira}, {Stott}, {Suchyta}, {Swanson}, {Tarle},
  {Thomas}, {Tucker}, {Uddin}, {Viana}, {Vikram}, {Walker}, {Zhang}, \& {DES
  Collaboration}}]{2016ApJS..224....1R}
{Rykoff}, E.~S., {Rozo}, E., {Hollowood}, D., {et~al.} 2016,
  \href{http://dx.doi.org/10.3847/0067-0049/224/1/1}{\JournalTitle{\apjs}, 224,
  1}

\bibitem[{{Sagunski} {et~al.}(2021){Sagunski}, {Gad-Nasr}, {Colquhoun},
  {Robertson}, \& {Tulin}}]{2020arXiv200612515S}
{Sagunski}, L., {Gad-Nasr}, S., {Colquhoun}, B., {Robertson}, A., \& {Tulin},
  S. 2021,
  \href{http://dx.doi.org/10.1088/1475-7516/2021/01/024}{\JournalTitle{\jcap},
  2021, 024}

\bibitem[{{Sameie} {et~al.}(2021){Sameie}, {Boylan-Kolchin}, {Sanderson},
  {Vargya}, {Hopkins}, {Wetzel}, {Bullock}, \& {Graus}}]{2021arXiv210212480S}
{Sameie}, O., {Boylan-Kolchin}, M., {Sanderson}, R., {et~al.} 2021,
  \JournalTitle{arXiv e-prints}, arXiv:2102.12480

\bibitem[{{Sameie} {et~al.}(2018){Sameie}, {Creasey}, {Yu}, {Sales},
  {Vogelsberger}, \& {Zavala}}]{2018MNRAS.479..359S}
{Sameie}, O., {Creasey}, P., {Yu}, H.-B., {et~al.} 2018,
  \href{http://dx.doi.org/10.1093/mnras/sty1516}{\JournalTitle{\mnras}, 479,
  359}

\bibitem[{{Sand} {et~al.}(2004){Sand}, {Treu}, {Smith}, \&
  {Ellis}}]{2004ApJ...604...88S}
{Sand}, D.~J., {Treu}, T., {Smith}, G.~P., \& {Ellis}, R.~S. 2004,
  \href{http://dx.doi.org/10.1086/382146}{\JournalTitle{\apj}, 604, 88}

\bibitem[{{Sawala} {et~al.}(2016){Sawala}, {Frenk}, {Fattahi}, {Navarro},
  {Bower}, {Crain}, {Dalla Vecchia}, {Furlong}, {Helly}, {Jenkins}, {Oman},
  {Schaller}, {Schaye}, {Theuns}, {Trayford}, \& {White}}]{2016MNRAS.457.1931S}
{Sawala}, T., {Frenk}, C.~S., {Fattahi}, A., {et~al.} 2016,
  \href{http://dx.doi.org/10.1093/mnras/stw145}{\JournalTitle{\mnras}, 457,
  1931}

\bibitem[{{Schaller} {et~al.}(2015){Schaller}, {Frenk}, {Bower}, {Theuns},
  {Jenkins}, {Schaye}, {Crain}, {Furlong}, {Dalla Vecchia}, \&
  {McCarthy}}]{2015MNRAS.451.1247S}
{Schaller}, M., {Frenk}, C.~S., {Bower}, R.~G., {et~al.} 2015,
  \href{http://dx.doi.org/10.1093/mnras/stv1067}{\JournalTitle{\mnras}, 451,
  1247}

\bibitem[{{Schneider}(2005)}]{2005astro.ph..9252S}
{Schneider}, P. 2005, \JournalTitle{arXiv e-prints}, astro

\bibitem[{Shin {et~al.}(2021)}]{Shin:2021odx}
Shin, T., {et~al.} 2021, \href{http://arxiv.org/abs/2105.05914}{{\sffamily
  arXiv:2105.05914 [astro-ph.CO]}}

\bibitem[{{Shirasaki}(2015)}]{2015ApJ...799..188S}
{Shirasaki}, M. 2015,
  \href{http://dx.doi.org/10.1088/0004-637X/799/2/188}{\JournalTitle{\apj},
  799, 188}

\bibitem[{{Sif{\'o}n} {et~al.}(2018){Sif{\'o}n}, {Herbonnet}, {Hoekstra}, {van
  der Burg}, \& {Viola}}]{2018MNRAS.478.1244S}
{Sif{\'o}n}, C., {Herbonnet}, R., {Hoekstra}, H., {van der Burg}, R. F.~J., \&
  {Viola}, M. 2018,
  \href{http://dx.doi.org/10.1093/mnras/sty1161}{\JournalTitle{\mnras}, 478,
  1244}

\bibitem[{{Sif{\'o}n} {et~al.}(2015){Sif{\'o}n}, {Cacciato}, {Hoekstra},
  {Brouwer}, {van Uitert}, {Viola}, {Baldry}, {Brough}, {Brown}, {Choi},
  {Driver}, {Erben}, {Grado}, {Heymans}, {Hildebrandt}, {Joachimi}, {de Jong},
  {Kuijken}, {McFarland}, {Miller}, {Nakajima}, {Napolitano}, {Norberg},
  {Robotham}, {Schneider}, \& {Verdoes Kleijn}}]{2015MNRAS.454.3938S}
{Sif{\'o}n}, C., {Cacciato}, M., {Hoekstra}, H., {et~al.} 2015,
  \href{http://dx.doi.org/10.1093/mnras/stv2051}{\JournalTitle{\mnras}, 454,
  3938}

\bibitem[{{Somerville}(2002)}]{2002ApJ...572L..23S}
{Somerville}, R.~S. 2002,
  \href{http://dx.doi.org/10.1086/341444}{\JournalTitle{\apjl}, 572, L23}

\bibitem[{{Spergel} \& {Steinhardt}(2000)}]{Spergel9909386}
{Spergel}, D.~N., \& {Steinhardt}, P.~J. 2000,
  \href{http://dx.doi.org/10.1103/PhysRevLett.84.3760}{\JournalTitle{\prl}, 84,
  3760}

\bibitem[{{Springel}(2005)}]{2005MNRAS.364.1105S}
{Springel}, V. 2005,
  \href{http://dx.doi.org/10.1111/j.1365-2966.2005.09655.x}{\JournalTitle{\mnras},
  364, 1105}

\bibitem[{{Springel} {et~al.}(2001){Springel}, {White}, {Tormen}, \&
  {Kauffmann}}]{2001MNRAS.328..726S}
{Springel}, V., {White}, S. D.~M., {Tormen}, G., \& {Kauffmann}, G. 2001,
  \href{http://dx.doi.org/10.1046/j.1365-8711.2001.04912.x}{\JournalTitle{\mnras},
  328, 726}

\bibitem[{Tanoglidis {et~al.}(2021)}]{Tanoglidis:2020wdy}
Tanoglidis, D., {et~al.} 2021,
  \href{http://dx.doi.org/10.3847/1538-4365/abca89}{\JournalTitle{Astrophys. J.
  Suppl.}, 252, 18}

\bibitem[{{Tollet} {et~al.}(2017){Tollet}, {Cattaneo}, {Mamon}, {Moutard}, \&
  {van den Bosch}}]{2017MNRAS.471.4170T}
{Tollet}, {\'E}., {Cattaneo}, A., {Mamon}, G.~A., {Moutard}, T., \& {van den
  Bosch}, F.~C. 2017,
  \href{http://dx.doi.org/10.1093/mnras/stx1840}{\JournalTitle{\mnras}, 471,
  4170}

\bibitem[{{Tulin} \& {Yu}(2018)}]{Tulin170502358}
{Tulin}, S., \& {Yu}, H.-B. 2018,
  \href{http://dx.doi.org/10.1016/j.physrep.2017.11.004}{\JournalTitle{\physrep},
  730, 1}

\bibitem[{{van den Bosch} \& {Ogiya}(2018)}]{VandenBosch180105427}
{van den Bosch}, F.~C., \& {Ogiya}, G. 2018,
  \href{http://dx.doi.org/10.1093/mnras/sty084}{\JournalTitle{\mnras}, 475,
  4066}

\bibitem[{{van den Bosch} {et~al.}(2018){van den Bosch}, {Ogiya}, {Hahn}, \&
  {Burkert}}]{VandenBosch171105276}
{van den Bosch}, F.~C., {Ogiya}, G., {Hahn}, O., \& {Burkert}, A. 2018,
  \href{http://dx.doi.org/10.1093/mnras/stx2956}{\JournalTitle{\mnras}, 474,
  3043}

\bibitem[{{van Dokkum} {et~al.}(2015){van Dokkum}, {Abraham}, {Merritt},
  {Zhang}, {Geha}, \& {Conroy}}]{2015ApJ...798L..45V}
{van Dokkum}, P.~G., {Abraham}, R., {Merritt}, A., {et~al.} 2015,
  \href{http://dx.doi.org/10.1088/2041-8205/798/2/L45}{\JournalTitle{\apjl},
  798, L45}

\bibitem[{{Vogelsberger} {et~al.}(2016){Vogelsberger}, {Zavala}, {Cyr-Racine},
  {Pfrommer}, {Bringmann}, \& {Sigurdson}}]{2016MNRAS.460.1399V}
{Vogelsberger}, M., {Zavala}, J., {Cyr-Racine}, F.-Y., {et~al.} 2016,
  \href{http://dx.doi.org/10.1093/mnras/stw1076}{\JournalTitle{\mnras}, 460,
  1399}

\bibitem[{{Wang} {et~al.}(2006){Wang}, {Li}, {Kauffmann}, \& {De
  Lucia}}]{2006MNRAS.371..537W}
{Wang}, L., {Li}, C., {Kauffmann}, G., \& {De Lucia}, G. 2006,
  \href{http://dx.doi.org/10.1111/j.1365-2966.2006.10669.x}{\JournalTitle{\mnras},
  371, 537}

\bibitem[{{Wechsler} \& {Tinker}(2018)}]{2018ARA&A..56..435W}
{Wechsler}, R.~H., \& {Tinker}, J.~L. 2018,
  \href{http://dx.doi.org/10.1146/annurev-astro-081817-051756}{\JournalTitle{\araa},
  56, 435}

\bibitem[{{Williamson} {et~al.}(2011){Williamson}, {Benson}, {High}, {Vand
  erlinde}, {Ade}, {Aird}, {Andersson}, {Armstrong}, {Ashby}, {Bautz}, {Bazin},
  {Bertin}, {Bleem}, {Bonamente}, {Brodwin}, {Carlstrom}, {Chang}, {Chapman},
  {Clocchiatti}, {Crawford}, {Crites}, {de Haan}, {Desai}, {Dobbs}, {Dudley},
  {Fazio}, {Foley}, {Forman}, {Garmire}, {George}, {Gladders}, {Gonzalez},
  {Halverson}, {Holder}, {Holzapfel}, {Hoover}, {Hrubes}, {Jones}, {Joy},
  {Keisler}, {Knox}, {Lee}, {Leitch}, {Lueker}, {Luong-Van}, {Marrone},
  {McMahon}, {Mehl}, {Meyer}, {Mohr}, {Montroy}, {Murray}, {Padin}, {Plagge},
  {Pryke}, {Reichardt}, {Rest}, {Ruel}, {Ruhl}, {Saliwanchik}, {Saro},
  {Schaffer}, {Shaw}, {Shirokoff}, {Song}, {Spieler}, {Stalder}, {Stanford},
  {Staniszewski}, {Stark}, {Story}, {Stubbs}, {Vieira}, {Vikhlinin}, \&
  {Zenteno}}]{Williamson11}
{Williamson}, R., {Benson}, B.~A., {High}, F.~W., {et~al.} 2011,
  \href{http://dx.doi.org/10.1088/0004-637X/738/2/139}{\JournalTitle{\apj},
  738, 139}

\bibitem[{{Zhang} {et~al.}(2019){Zhang}, {Jeltema}, {Hollowood}, {Everett},
  {Rozo}, {Farahi}, {Bermeo}, {Bhargava}, {Giles}, {Romer}, {Wilkinson},
  {Rykoff}, {Mantz}, {Diehl}, {Evrard}, {Stern}, {Gruen}, {von der Linden},
  {Splettstoesser}, {Chen}, {Costanzi}, {Allen}, {Collins}, {Hilton}, {Klein},
  {Mann}, {Manolopoulou}, {Morris}, {Mayers}, {Sahlen}, {Stott}, {Vergara
  Cervantes}, {Viana}, {Wechsler}, {Allam}, {Avila}, {Bechtol}, {Bertin},
  {Brooks}, {Burke}, {Carnero Rosell}, {Carrasco Kind}, {Carretero},
  {Castander}, {da Costa}, {De Vicente}, {Desai}, {Dietrich}, {Doel},
  {Flaugher}, {Fosalba}, {Frieman}, {Garc{\'\i}a-Bellido}, {Gaztanaga},
  {Gruendl}, {Gschwend}, {Gutierrez}, {Hartley}, {Honscheid}, {Hoyle},
  {Krause}, {Kuehn}, {Kuropatkin}, {Lima}, {Maia}, {Marshall}, {Melchior},
  {Menanteau}, {Miller}, {Miquel}, {Ogando}, {Plazas}, {Sanchez}, {Scarpine},
  {Schindler}, {Serrano}, {Sevilla-Noarbe}, {Smith}, {Soares-Santos},
  {Suchyta}, {Swanson}, {Tarle}, {Thomas}, {Tucker}, {Vikram}, {Wester}, \&
  {DES Collaboration}}]{2019MNRAS.487.2578Z}
{Zhang}, Y., {Jeltema}, T., {Hollowood}, D.~L., {et~al.} 2019,
  \href{http://dx.doi.org/10.1093/mnras/stz1361}{\JournalTitle{\mnras}, 487,
  2578}

\end{thebibliography}


\appendix

\section{Convergence Tests} \label{conv_test}

\begin{figure} \label{Figure:conv_test}
\centering
\includegraphics[scale=0.35,trim={0 0.1cm 0 0},clip]{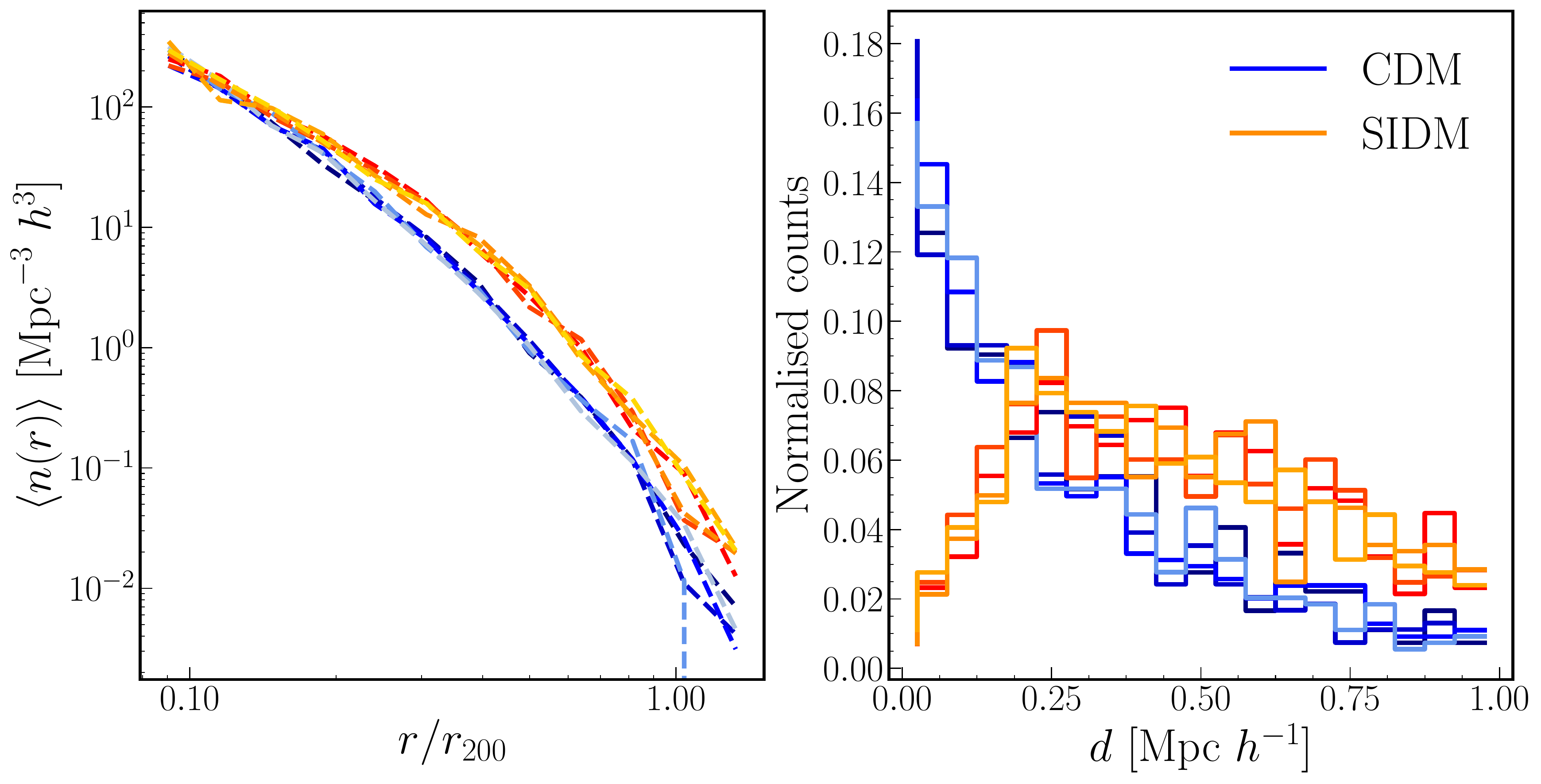} 
\caption{\textit{Left:} The variation in the radial distributions, $\langle n \rangle$ of disrupted subhalos (\vpeak{} >136.5 \kms{}) when the tracer particle is varied based on boundedness. The \textit{dashed} curves of five different shades for SIDM (\textit{orange}) and CDM (\textit{blue}) correspond to the first five most bound particles. \textit{Right:} The distributions of the distances at $z=0$ between the most bound particle and each of the $n^{\rm th}$ most bound tracers.}
\end{figure}

Here we test the validity of our choice to track the disrupted subhalos using the MBP. We study the convergence properties of the radial distribution of the disrupted subhalos by varying the choice of the tracer from the most bound to the $n^{\rm th}$ bound particle in the subhalo at the time it achieves its peak velocity. For this purpose, the MBP aside, we record the $z=0$ positions of the $n$th most bound particles, i.e., the particle with $n$th least total energy $E$, where $n \in \{2,3,4,5\}$. We evaluate the 3D distances between each of four types of tracers and the fiducial most bound tracer plot their distributions in Fig.\ \ref{Figure:conv_test}. The radial distribution of the tracers is also shown alongside the histograms of their mutual separations in the same figure.

We find that the radial distribution of the disrupted subhalos is well converged both in SIDM and CDM, and does not change significantly when the tracer is shifted among the first few most bound particles. This implies that we will have a robust prediction for the orphan fraction as a function of radius in our analysis using the MBPs. However, we note that the distributions of the mutual separation vary significantly between the two DM models, wherein majoritiy of the tracers in CDM generally end up close to each other, $< 0.25$ \mpch{} with very few at distances $> 0.5$ \mpch{}. On the other hand, the mutual separation of the tracers in SIDM follows a skewed distribution peaking at $0.25$ \mpch{}. This implies that as a result of scatterings, the particles near the potential minima of a disrupted subhalo can be scattered to large distances from the center of the subhalo, and it is non-trivial to assign galaxy positions to these disrupted systems based solely on their MBPs. 

We don't expect the latter results to affect our predictions for weak lensing and our inferences through the paper, as we primarily rely on trying to rule out a model of CDM + orphans. However, as the profile around MBPs in SIDM is mostly flat, particularly for disrupted subhalos in the inner regions of the cluster, our work incorporates the maximal disruption scenario in SIDM. Note that we do notice the faint remnants of the disrupted cores in SIDM in the cluster outskirts, which demonstrates the evidence of structure around the MBP. Modeling the galaxy-halo connection for disrupted systems like these will be able to provide a clearer picture.

\section{Stellar mass contribution} \label{sec:baryons}

In this section we describe our approach to address the contribution from the stellar mass of the central galaxy to \esd{}$(R)$ around subhalos. Whereas our \esd{} profiles are stacked in bins of the subhalos' \vpeak{}, the observed profiles will be around galaxies selected and binned according to their luminosities. Here we apply an empirical relation employed in \citet{2018MNRAS.477..359C} to assign stellar masses to the subhalos using the values of their \vpeak{}. However we note that this relation is agnostic to the halos being centrals or subhalos and the galaxies contained in them can evolve even after they are accreted onto their host. For the sake of simplicity, we will neglect the evolution of these satellite galaxies initially, and start with a non-evolving, i.e., redshift independent model of matching $M_{\star}-V_{\rm peak}$,
\begin{equation}
M_{\star}(V_{\rm peak}) = 2 M_0 \left(\frac{V_{\rm peak}}{V_0}\right) \left[ \left(\frac{V_{\rm peak}}{V_0}\right)^{\alpha} + \left(\frac{V_{\rm peak}}{V_0}\right)^{\beta}  \right]^{-1}    
\end{equation}
According to \citet{2018MNRAS.477..359C} a good fit is provided by, ${\rm log}(M_0)=9.95\pm0.01, {\rm log}(V_0) =2.177\pm0.005,\alpha =-5.9\pm0.1$ and $\beta=-0.25\pm0.02$ . The contribution of stars to the stacked subhalo \esd{} profile $\Delta \Sigma_{\star}$ is obtained by assuming that the distribution of the stars can be treated as a centrally located point with respect to the extent of the dark matter halo. 
\begin{equation}
\Delta \Sigma_{\star}(R) = \dfrac{M_{\star}}{\pi R^2} 
\end{equation}
The mean value of $M_{\star}$ is calculated for the relevant bin of \rsat{} and \vpeak{} and the $\Delta \Sigma_{\star}$ thus derived is added to the existing dark matter only \esd{} profile.
\begin{equation}
\Delta \Sigma (R,R_{\rm sub}) = \Delta \Sigma_{\rm sub} (R) + \Delta \Sigma_{\rm host} (R,R_{\rm sub}) + \Delta \Sigma_{\star}(R)
\end{equation}


\section{Miscentering} \label{sec:miscent}

The positions of the centers of the simulated halos are determined in {\sc Rockstar} as the mean position of a confined set of particles around the density peak of the halo \citep{2013ApJ...762..109B}. We artificially miscenter the \esd{} profiles we obtain from the simulations in a Monte Carlo fashion to ascertain its effect on the mock lensing observable analysis. 

We apply the method of \citet{2017MNRAS.469.4899M}, assuming that a fraction $f_{mis}$ of the cluster centers are miscentered and the stacked \esd{} profile around their subhalos is $\Delta \Sigma_{mis}$. Similarly the fraction of clusters that are well centered have a stacked subhalo \esd{} profile $\Delta \Sigma_0$,
\begin{equation}
\Delta \Sigma = (1 - f_{\rm mis}) \, \Delta \Sigma_0 + f_{\rm mis} \, \Delta \Sigma_{\rm mis}
\end{equation}
The miscentered clusters are chosen by randomly sampling a fraction $f_{\rm mis}$ of all the simulated clusters. The profile of each surviving and disrupted subhalo belonging to each of them is miscentered by reassigning a new value of \rsat{} to them. We obtain the miscentered profiles by stacking in bins of $R_{\rm sub, mis}$, the projected distances between the host center and subhalo centers that have been modified by using the cosine law,
\begin{equation}
R_{\rm sub,mis} = \sqrt{R^2_{\rm sub} + R^2_{\rm off} + 2 R_{\rm sub} R_{\rm off} {\rm cos} \theta}
\end{equation}
Here we assume that the uncertainties on the position of the subhalo centers are negligible compared to the miscentering offset of their hosts. A value of the offset radius $R_{\rm off}$ is sampled from a Rayleigh distribution \citep{2007arXiv0709.1159J} that is parameterized with $\sigma_{\rm off}$ which represents the mode of the distribution. 
\begin{equation}
P(R_{\rm off}) = \frac{R_{\rm off}}{\sigma_{\rm off}} \exp \bigg(- \frac{R^2_{\rm off}}{2\sigma^2_{\rm off}} \bigg)
\end{equation}
We choose a value of $\sigma_{\rm off}=0.2$ \mpch{} and $f_{\rm mis}=0.22$. Likewise, for each subhalo, ${\rm cos} \theta$ is drawn from a uniform distribution in the interval $(-1,1)$

\begin{figure}
\centering
\includegraphics[scale=0.35,trim={0 0.3cm 0 0.2cm},clip]{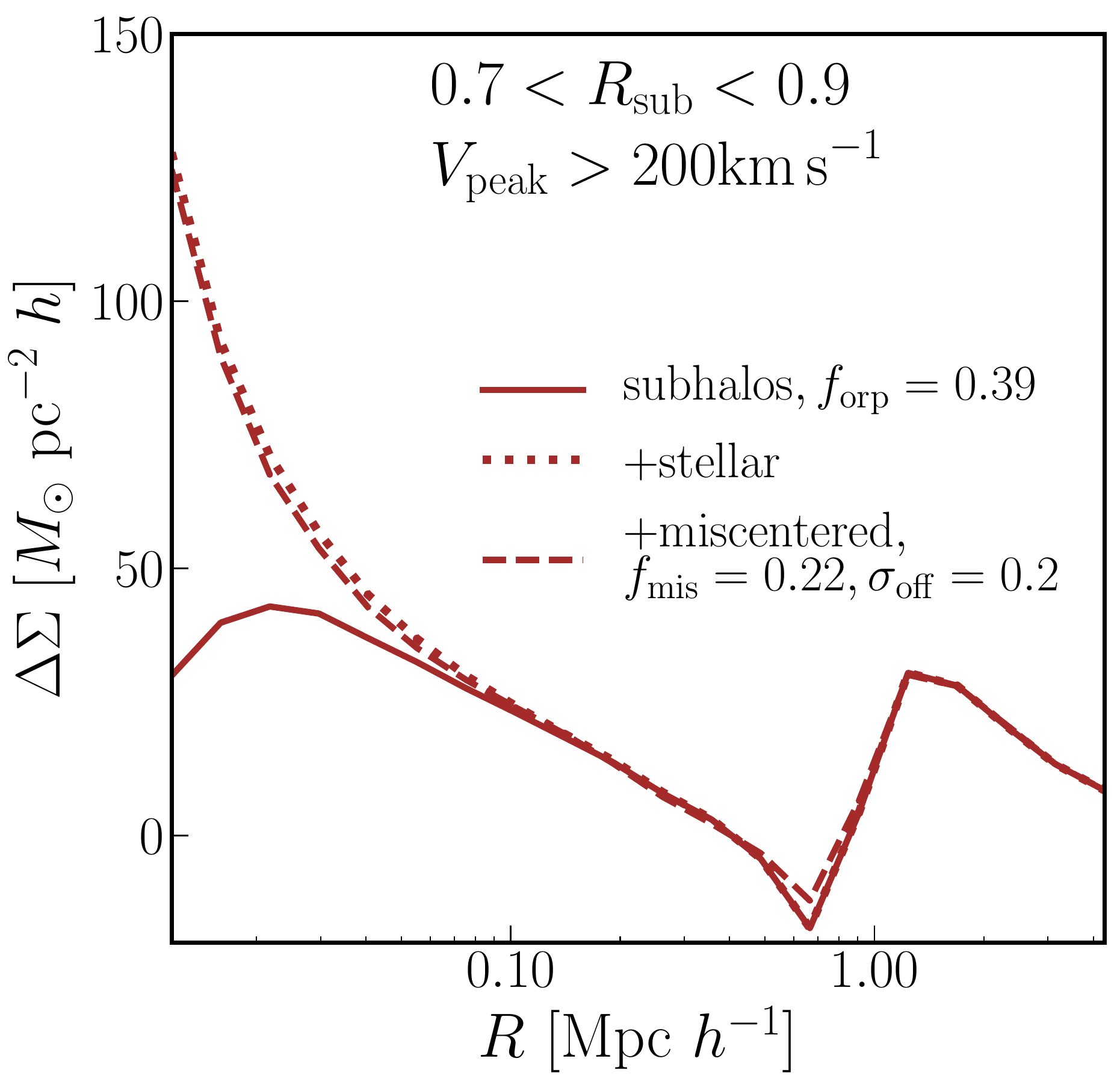}
\caption{The dark matter only profile with an orphan fraction $f_{\rm orp}=0.38$ is depicted by the \textit{solid} line and the addition of the stellar component is shown with a \textit{dotted} line. When the hosts are miscentered using $f_{\rm orp}=0.22$ and $\sigma_{\rm off}=0.2$ \mpch{} followed by stacking, the result is the \textit{dashed} line. }
\label{Figure:esd_syst}
\end{figure}



\end{document}